\documentclass[reprint,superscriptaddress,showpacs,showkeys,amsmath,amssymb,floatfix,aps,pra,longbibliography,groupedaddress]{revtex4-1}

\usepackage{graphicx}% Include figure files
\usepackage{bm}% bold math

\usepackage[dvipsnames]{xcolor}
\usepackage{physics}
\usepackage{placeins}
\usepackage{graphics}
\usepackage{color}
\usepackage{hyperref}
\usepackage{multirow}
\usepackage{blindtext}
\usepackage{gensymb}
\usepackage{textcomp}
\begin{document}

%\preprint{APS/123-QED}

\title{Twist-angle dependent proximity induced spin-orbit coupling\\ in graphene/topological insulator heterostructures}

\author{Thomas Naimer}
\email{thomas.naimer@physik.uni-regensburg.de}
%\affiliation{Institute for Theoretical Physics, University of Regensburg, 93040 Regensburg, Germany}
\author{Jaroslav Fabian}
\affiliation{Institute for Theoretical Physics, University of Regensburg, 93040 Regensburg, Germany}

\begin{abstract}
The proximity-induced spin-orbit coupling (SOC) in heterostructures of twisted graphene and topological insulators (TIs) Bi$_2$Se$_3$ and Bi$_2$Te$_3$ is investigated from first principles. To build commensurate supercells, we strain graphene and correct thus resulting band offsets by applying a transverse electric field. We then fit the low-energy electronic spectrum to an effective Hamiltonian that comprises orbital and spin-orbit terms. For twist angles 0$\degree\leq\Theta \lessapprox 20\degree$, we find the dominant spin-orbit couplings to be of the  valley-Zeeman and Rashba types, both a few meV strong. We also observe a sign change in the induced valley-Zeeman SOC at $\Theta\approx 10\degree$. Additionally, the in-plane spin structure resulting from the Rashba SOC acquires a non-zero radial component, except at $0\degree$ or $30\degree$. At $30\degree$  the graphene Dirac cone interacts directly with the TI surface state. We therefore explore this twist angle in more detail, studying the effects of gating, TI thicknesses, and lateral shifts on the SOC parameters. % and different fitting Hamiltonians 
We find, in agreement with previous results, the emergence of the proximitized  Kane-Mele SOC, with a change in sign possible by electrically tuning the Dirac cone within the TI bulk band gap. 
%We further discuss the reliability of the supercell approach with respect to atomic relaxation (rippling of graphene),the thickness of the TI and relative lateral shifts of the atomic layers.

\end{abstract}

\maketitle

%------------------------------------------------------------
\section{Introduction}
%------------------------------------------------------------
Its long spin-relaxation times~\cite{Singh2016:APL,Drogeler2016:NL} and high electronic mobility~\cite{Bolotin2008:SuspendedGrapheneHighmobility1,Du2008:SuspendedGrapheneHighmobility2} make graphene an excellent material for spin transport. However, to enable the manipulation of the spins in graphene and therefore truly unlock its full potential as a platform for spintronics, enhancing and controling graphene's spin-orbit coupling (SOC) is necessary.  Additionally, this control can enable the creation of multiple topological states~\cite{Kane2005:PRL,Kane2005:PRL2,Qiao2010:PRB,Ren2016:RPP,Frank2018:PRL,Hogl2020:PRL}. Different features can be achieved by adding different flavours of SOC to the graphene: While Kane-Mele type SOC will lead to the formation of a Quantum Spin Hall effect (QSHE)~\cite{Kane2005:PRL,Kane2005:PRL2}, valley-Zeeman type SOC is needed to  create spin-orbit valves~\cite{Khoo2017:NL,Gmitra2017:PRL,Island2019:SOValve2,Tiwari2021:SOValve3,Amann2021:arxiv} using giant spin-relaxation anisotropy~\cite{Zihlmann2018:PRB,Cummings2017:PRL,Ghiasi2017:NL}. Furthermore, Rashba type SOC will produce a Rashba Edelstein effect (REE, charge-to-spin conversion) or even the recently discussed unconventional Rashba Edelstein effect (UREE, collinear charge-to-spin conversion)~\cite{Zhao2022:expradialrashba,Ingla_Aynes2022:expradialrashba2,Peterfalvi2022:radialrashba,Veneri22:PRB:radialrashba2,Lee22:PRB:radialrashba3}.

A very successful way of realizing such systems has been building van der Waals heterostructures~\cite{Geim2013:Nat,Duong2017:ACS}, in which---due to proximity effects---electronic properties of one two-dimensional material can be transferred to another. To induce SOC in graphene, using transition-metal dichalcogenides (TMDCs such as WSe$_2$) to form graphene/TMDC heterostructures~\cite{Gmitra2015:TMDCgraphene1,Gmitra2016:TMDCgraphene2,Frank2017:PRB,Wang2015:NC,Avsar2014:NC} has proven to be a viable route. Using thin layers of three-dimensional (3D) topological insulators (TIs) like Bi$_2$Te$_3$ or Bi$_2$Se$_3$  can introduce even larger SOC~\cite{Song2018:TIGRheteroDFT,Zollner2019b:PRB}. In experiment, such graphene/TI heterostructures can be fabricated using either exfoliation techniques~\cite{Jafarpisheh2018:PRB,Khokhriakov2018:TIGRheteroexp,Steinberg2015:PRB,Zalic2017:PRB,Rajput16:acs} or techniques like chemical vapor deposition (CVD)~\cite{Dang2010:NL,Lee15:acs,Kiemle22:acs,Song2010:APL}. While the former case should result in incommensurate structures with random twist angle, the latter will produce commensurate structures with mostly fixed twist angle $\Theta$ ($\Theta=0\degree$ or $\Theta=30\degree$)~\cite{Kiemle22:acs,Dang2010:NL} between the two layers. Similar to graphene/TMDC heterostructures, the twist angle in graphene/TI heterostructures will also play a significant role for the proximity SOC that graphene obtains.

Apart from simple low energy models~\cite{Zhang14:PRB:TIonlyTB,DeBeule17:PRB:TIonlyTB}, a good approach for theoretically describing such heterostructures is the \emph{ab initio} approach of density functional theory (DFT). Previous papers employing DFT for graphene/TI structures~\cite{Cao16:2DM,Jin2013:PRB,Khokhriakov2018:TIGRheteroexp,Lee15:acs,Lin17:NL,Liu13:PRB:TIGrDFT,Popov14:PRB,Rajput16:acs} focus mainly on the 30\degree~supercell, while some also explore the 0\degree~case~\cite{Zollner2019b:PRB,Song2018:TIGRheteroDFT}. However, intermediate twist angles have not been yet considered, although it is at those angles where symmetry allows for a radial in-plane spin structure and therefore the UREE to arise.

In this manuscript we make a comprehensive DFT study of the proximity SOC of graphene/Bi$_2$Te$_3$ and graphene/Bi$_2$Se$_3$ heterostructures for a set of commensurate supercells with different twist angles and strains. To increase the comparability between the supercells, we correct the band offsets (arising due to strain) by applying a transverse electric field, 
as already discussed in Ref.~\cite{Naimer2021:paper1} for graphene/TMDC. We find a combination of valley-Zeeman and Rashba type SOC to be induced for all cases except 30\degree. Additionally, we observe a large twist-angle dependency of the Rashba phase angle, indicating the possibility of a purely radial in-plane spin structure (90\degree~phase angle) and UREE. 
A purely radial spin structure was already proposed to occur in graphene/TMDC heterostructures by a tight binding model~\cite{Peterfalvi2022:radialrashba}. However, DFT calculations~\cite{Naimer2021:paper1,Lee22:PRB:radialrashba3} on such graphene/TMDC heterostructures are in disagreement with that prediction, finding a maximal phase angle of $\pm30\degree$. Furthermore, we discuss the band structure of the 30\degree~supercell in more detail. Here, our results (i.e. the delicate spin structure and the unique appearance of Kane-Mele SOC) are in good agreement with Ref.~\cite{Song2018:TIGRheteroDFT}. However, we additionally explore the effect of an external transverse electric field and find that it can change the sign of the Kane-Mele SOC.
%{\color{red} Add more to this teaser, say also more what has been known, especially from Song et al, and how we go beyond that. Give referenes to Gr/TMDC papers on non-radial Rashba, say how much it deviates here. etc. Check. }

The manuscript is organized as follows. In Sec.~\ref{Sec:methods} we introduce the methodology, the supercell structures and the procedure we apply to adjust for strain induced changes in the band offsets. Sec.~\ref{Sec:Umklapp} shows how--- following Ref.~\cite{Koshino2015:TwistTBBasic}---the graphene Dirac cone couples to different parts of the TI 1.BZ for different twist angles and strains. The model Hamiltonian used to fit the results is presented in Sec.~\ref{Sec:effHam}. The fitting results are shown in Sec.~\ref{Sec:Res}. We present twist angles $0\degree\leq\Theta\lessapprox 20\degree$ and  $\Theta=30\degree$  in Subsec.~\ref{Subsec:ResI} and Subsec.~\ref{Subsec:ResII} respectively. In App.~\ref{App:comp} computational details are given. In App.~\ref{App:finestruc} and App.~\ref{App:Aaronfitting} we discuss details of the 30\degree~supercell band structure: App.~\ref{App:finestruc} explains what we call 'type 1' and 'type 2' band pairs in the main manuscript and App.~\ref{App:Aaronfitting} discusses an alternative fitting Hamiltonian. In App.~\ref{App:shift} and App.~\ref{App:QL} the effects of varying lateral shifts and varying TI thicknesses are discussed.

%------------------------------------------------------------
\section{Methods}
\label{Sec:methods}
%------------------------------------------------------------
The hexagonal unit cell of the TIs is described by the two lattice parameters $a$ and $c$ as well as the atomic constants $u$ and $v$ (all geometry parameters for Bi$_2$Se$_3$ and Bi$_2$Te$_3$ are listed in Tab.~\ref{Tab:latconsts}). While such a unit cell contains 3 quintuple layers (QLs), we use, unless specified otherwise, only one QL to reduce computational effort.  Due to the short range of proximity effects, the effect of additional QLs on the graphene is almost exclusively via the change of the TI surface state. Since heterostructures with $\Theta\neq30\degree$ explicitly do not couple to the surface state (see Sec.~\ref{Sec:Umklapp}), and in accordance with the results of Ref.~\cite{Zollner2019b:PRB}, we deem the 1QL cases to be representative for $0<\Theta\lessapprox20\degree$. We discuss 3QL cases in App.~\ref{App:QL} and in Subsec.~\ref{Subsec:ResII}, in connection with $\Theta = 30\degree$.
%{\color{red} Discuss already here why we think that a single QL should be representative except for 30. Check, but why did you delete my sentence, where I talk about exactly that, only for me to rewrite it now?}

%-----------------------------------------------------------------
    \begin{table}[htb]
    \caption{Lattice constants and atomic constants of unstrained primitive unit cells of graphene, Bi$_2$Te$_3$ and Bi$_2$Se$_3$~\citep{Nakajima1963:TIlatconst}. The structure of the TIs stays unchanged in graphene/TI supercells we use, while the graphene layers are strained by the factors $\epsilon$ listed in Tab.~\ref{Tab:Structures} to ensure commensurability.}\label{Tab:latconsts}
        \begin{ruledtabular}
        \begin{tabular}{c|cccc}

				& $a$[\AA] 	& $c$[\AA]& $u$[\AA]  &	$v$[\AA]\\
				\hline
 Graphene		&	2.46 	&  - 	&  - &-\\
 Bi$_2$Te$_3$		&	4.386	& 30.497 	& 0.4000$\cdot c$ 	& 0.2097$\cdot c$\\
Bi$_2$Se$_3$ 		&	4.143  	&28.636   	&0.4008$\cdot c$   &0.2117$\cdot c$\\

     	\end{tabular}
        \end{ruledtabular}

    \end{table}
%------------------------------------------------------------------------ 
We construct the supercells by implementing the coincidence lattice method~\citep{Koda2016:JPCC,Carr2020:NRM}, which is detailed in Ref.~\citep{Naimer2021:paper1}. We give integer attributes $(n,m)$ to a monolayer supercell. The lattice vectors $\mathbf{a}^S_{(n,m)}$ and $\mathbf{b}^S_{(n,m)}$  are defined as a linear combination of the primitive lattice vectors $\mathbf{a}$ and $\mathbf{b}$:
\begin{align}
\mathbf{a}^S_{(n,m)}&=n\cdot \mathbf{a}+m\cdot \mathbf{b} \\
\mathbf{b}^S_{(n,m)}&=-m\cdot \mathbf{a}+(n+m)\cdot \mathbf{b}.
\end{align}
By placing an ($n$,$m$) graphene supercell beneath an ($n'$,$m'$) TI supercell, we construct the graphene/TI heterostructure supercell, which then has a certain relative twist angle $\Theta$ depending on $n$, $m$, $n'$ and $m'$.

\begin{figure}[htbp]
\includegraphics[width=.99\linewidth]{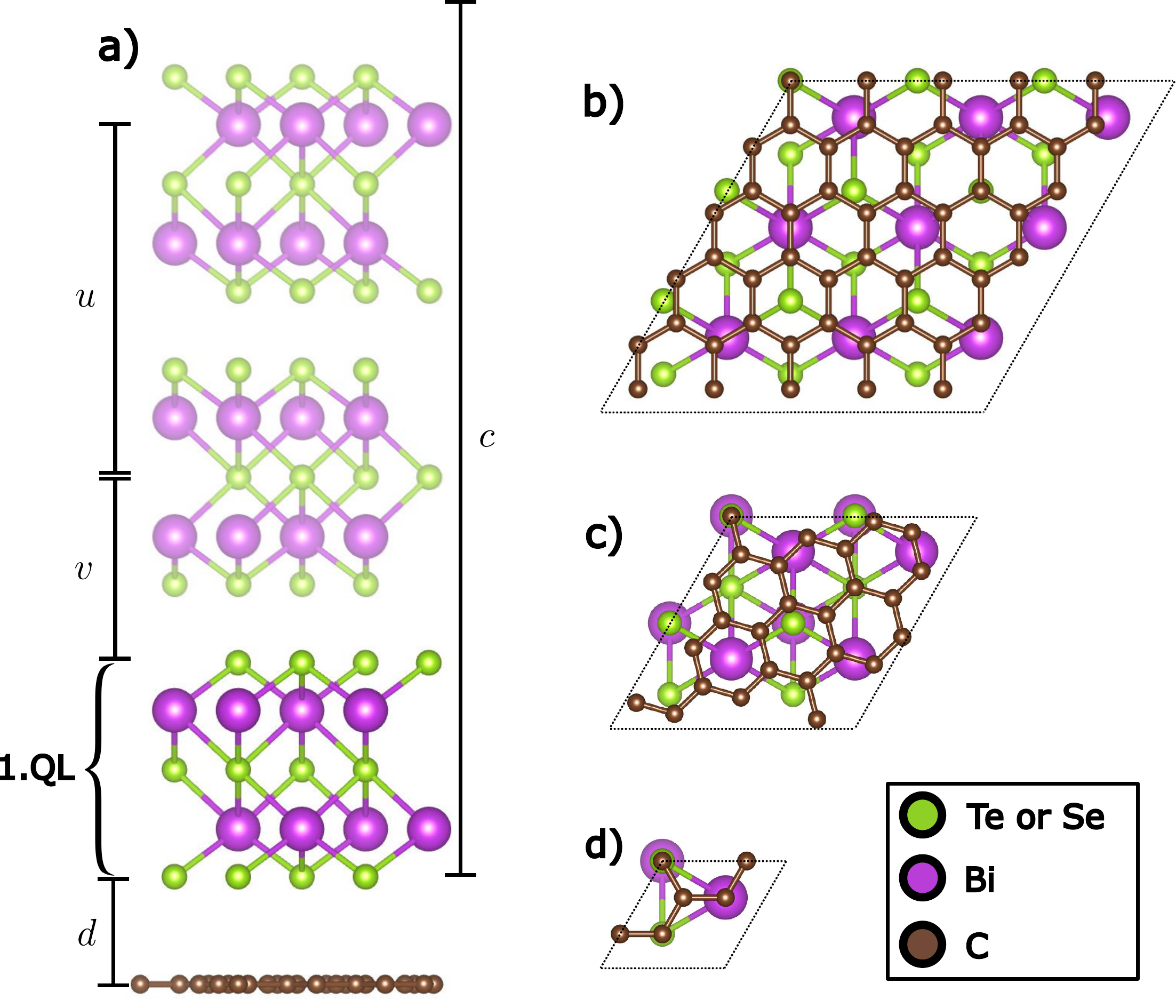} 
     \caption{ (a) Side view of the 13.9\degree~unit cell with indicated interlayer distance $d$, atomic constants $u$ and $v$ and lattice constant $c$. The upper two QLs are transparent, since mostly only the first QL is used in our calculations. (b), (c) and (d) Bottom view of the graphene/TI heterostructure supercells with twist angles $\Theta= 0\degree,13.9\degree$, and $30\degree$, respectively.}\label{Fig:Structures}
  \end{figure}

 If not specified otherwise, our supercells follow the convention of a 'Top' configuration. This means that at a corner of the supercell a carbon atom resides directly beneath a Te or Se atom, see Fig.~\ref{Fig:Structures}. Considering different configurations (see App.~\ref{App:shift}) we find that for large enough supercells
  the proximity SOC is rather insensitive to the changes of the atomic registry, similar to what is observed in graphene/TMDC heterostructures~\citep{Naimer2021:paper1,Gmitra2016:TMDCgraphene2}.

In order to obtain commensurate supercells for periodic DFT calculations, one of the layers (or both) need to be strained. We thus introduce the strain factor  $\epsilon$, which depends on the lattice constant of the TI and is therefore different for Bi$_2$Se$_3$ and Bi$_2$Te$_3$. Since the low energy Dirac spectrum of graphene is (apart from the renormalization of the Fermi velocity) rather robust against biaxial strain smaller than 20\%~\citep{Si2016:graphenestrain1,Choi2010:graphenestrain2}, we choose to leave the TI unstrained and strain graphene.

%------------------------------------------------------------------------

\begin{table}[]
\caption{\label{Tab:Structures} Structural information of the investigated graphene/TI heterostructures. 
   Listed are the supercell attributes $(n,m)$ of graphene and $(n',m')$ of the TI, the resulting twist angle $\Theta$ between the two monolayers and the strain $\epsilon^{\text{TI}}$ imposed on the graphene (which depends on the specific choice of TI). Additionally, we list the number of atoms ($N_{\text{at}}$) in the heterostructure for the cases with 1 and 3 Quintuple layers (QL) of TI. For completeness, we also list strains and $N_{\text{at}}$ corresponding to supercells, which were not investigated, in non-boldface. These supercells, which were not investigated, either had too much built-in strain, too many atoms or entailed computational difficulties (e.g., convergence problems).}
\begin{tabular}{c|cc|cc|cc}

$\Theta $ &$(n,m)$ & $(n',m')$& $\epsilon^{\text{Bi}_2\text{Te}_3}$ & $\epsilon^{\text{Bi}_2\text{Se}_3}$&$N_\text{at}$&$N_\text{at}$   \\
$[\degree]$&&&[\%]&[\%]&(1QL)&(3QL)\\
\hline

%0.0 	 &( 0 2 )	 &( 0 1 )      & -10.85 & -15.79        & 13  &\textcolor{gray}{23}\\
%0.0 	 &( 0 5 )	 &( 0 3 )      & 6.98 & 1.05        & 95 & \textcolor{gray}{185} \\ 
%4.3 	 &( 2 3 )	 &( 1 2 )      & 8.22 & 2.22        & 73 &143 \\ 
%4.7 	 &( 4 3 )	 &( 2 2 )      & 1.54 & -4.09        & 134  & \textcolor{gray}{254}\\
%8.9 	 &( 1 5 )	 &( 0 3 )      & -3.93 & \textcolor{gray}{-9.26}        & 107 & \textcolor{gray}{197} \\ 
%10.9 	 &( 2 1 )	 &( 1 1 )      & 16.72 & 10.25        & 29 &59 \\ 
%13.9 	 &( 1 3 )	 &( 0 2 )      & -1.1 & -6.58        & 46 &  86\\ 
%16.1 	 &( 3 1 )	 &( 1 1 )      & -14.35 & -19.1        &41 & \textcolor{gray}{71}  \\ 
%17.5 	 &( 3 2 )	 &( 1 2 )      & 8.22 & 2.22        & 73& \textcolor{gray}{143} \\
%19.1 	 &( 4 0 )	 &( 2 1 )      & 17.93 & 11.4        & 67 &  \textcolor{gray}{137}\\ 
%19.1 	 &( 5 0 )	 &( 2 1 )      & -5.66 & \textcolor{gray}{-10.88}        & 85 &  \textcolor{gray}{155}\\ 
%19.1     &( 2 4 )        &( 0 3 )      & 1.08 & -4.52     & 101  &\textcolor{gray}{191}  \\
%20.8     &( 4 3 )        &( 1 3 )      & \textcolor{gray}{5.68} & -0.17   &  139    & \textcolor{gray}{269} \\
%21.1     &( 5 1 )        &( 2 2 )      & \textcolor{gray}{10.93} & 4.78    &  122  &\textcolor{gray}{242}   \\
%21.8 	 &( 4 2 )	 &( 1 2 )      & -10.85 & \textcolor{gray}{-15.79}        & 91 & \textcolor{gray}{161} \\ 
%30.0 	 &( 1 1 )	 &( 0 1 )      & 2.94 & -2.77        & 11 & 21 \\ 
%30.0     &( 7 0 )        &( 2 2 )      & -11.77 & -16.66   &   158  & \textcolor{gray}{278} \\

0.0 	 &( 0 2 )	 &( 0 1 )      & \textbf{-10.85} &\textbf{-15.79}         & \textbf{13}  &{23}\\
0.0 	 &( 0 5 )	 &( 0 3 )      & \textbf{6.98} & \textbf{1.05}        & \textbf{95} & {185} \\ 
4.3 	 &( 2 3 )	 &( 1 2 )      & \textbf{8.22} & \textbf{2.22}       & \textbf{73} &\textbf{143} \\ 
4.7 	 &( 4 3 )	 &( 2 2 )      & \textbf{1.54} & \textbf{-4.09}        & \textbf{134}  & {254}\\
8.9 	 &( 1 5 )	 &( 0 3 )      & \textbf{-3.93} & -9.26              & \textbf{107} & {197} \\ 
10.9 	 &( 2 1 )	 &( 1 1 )      & \textbf{16.72} & \textbf{10.25}        & \textbf{29} &\textbf{95} \\ 
13.9 	 &( 1 3 )	 &( 0 2 )      & \textbf{-1.1} & \textbf{-6.58}        & \textbf{46} &  \textbf{86}\\ 
16.1 	 &( 3 1 )	 &( 1 1 )      & \textbf{-14.35} & \textbf{-19.1}        &\textbf{41} & {71}  \\ 
17.5 	 &( 3 2 )	 &( 1 2 )      & \textbf{8.22} & \textbf{2.22 }       & \textbf{73}& {143} \\
19.1 	 &( 4 0 )	 &( 2 1 )      & \textbf{17.93} & \textbf{11.4}        & \textbf{67} &  {137}\\ 
19.1 	 &( 5 0 )	 &( 2 1 )      & \textbf{-5.66} & -10.88                & \textbf{85} &  {155}\\ 
19.1     &( 2 4 )    &( 0 3 )      & \textbf{1.08} & \textbf{-4.52}     & \textbf{101}  &{191}  \\
20.8     &( 4 3 )    &( 1 3 )      & 5.68           & \textbf{-0.17}   &  \textbf{139}    & {269} \\
21.1     &( 5 1 )    &( 2 2 )      & 10.93          & \textbf{4.78}    &  \textbf{122}  &{242}   \\
21.8 	 &( 4 2 )	 &( 1 2 )      & \textbf{-10.85} & \textbf{-15.79}        & \textbf{91} & {161} \\ 
30.0 	 &( 1 1 )	 &( 0 1 )      & \textbf{2.94} & \textbf{-2.77}        & \textbf{11} & \textbf{21} \\ 
30.0     &( 7 0 )    &( 2 2 )      & \textbf{-11.77} & \textbf{-16.66 }  &  \textbf{158}   & {278} \\

\end{tabular}

\end{table}
Also, to focus on twist-angle effects, the used interlayer distance $d=3.5$~\AA~ separating the monolayers (see Fig.~\ref{Fig:Structures}) is the same for all studied supercells.
%Nevertheless, in App.~\ref{App:relax} the effects of structural relaxation and rippling on the proximity band structure at the Dirac cone are discussed. 
We do not perform structural relaxation calculations for our systems assuming that -- as discussed in \citep{Naimer2021:paper1} for TMDCs -- this will only lead to a modification of the staggered potential (due to rippling effects) and leave the SOC parameters largely unaffected.
To avoid interactions between periodic images in our slab geometry, we add a vacuum of 20~\AA. All graphene/TI heterostructures are set up using the {\tt atomic simulation environment (ASE)}~\citep{ASE} code.
The structural parameters of the heterostructures are collected in  Tab.~\ref{Tab:Structures} and some representative examples are visualized in Fig.~\ref{Fig:Structures}. 
    
\begin{figure}[htb]
     \includegraphics[width=.8\linewidth]{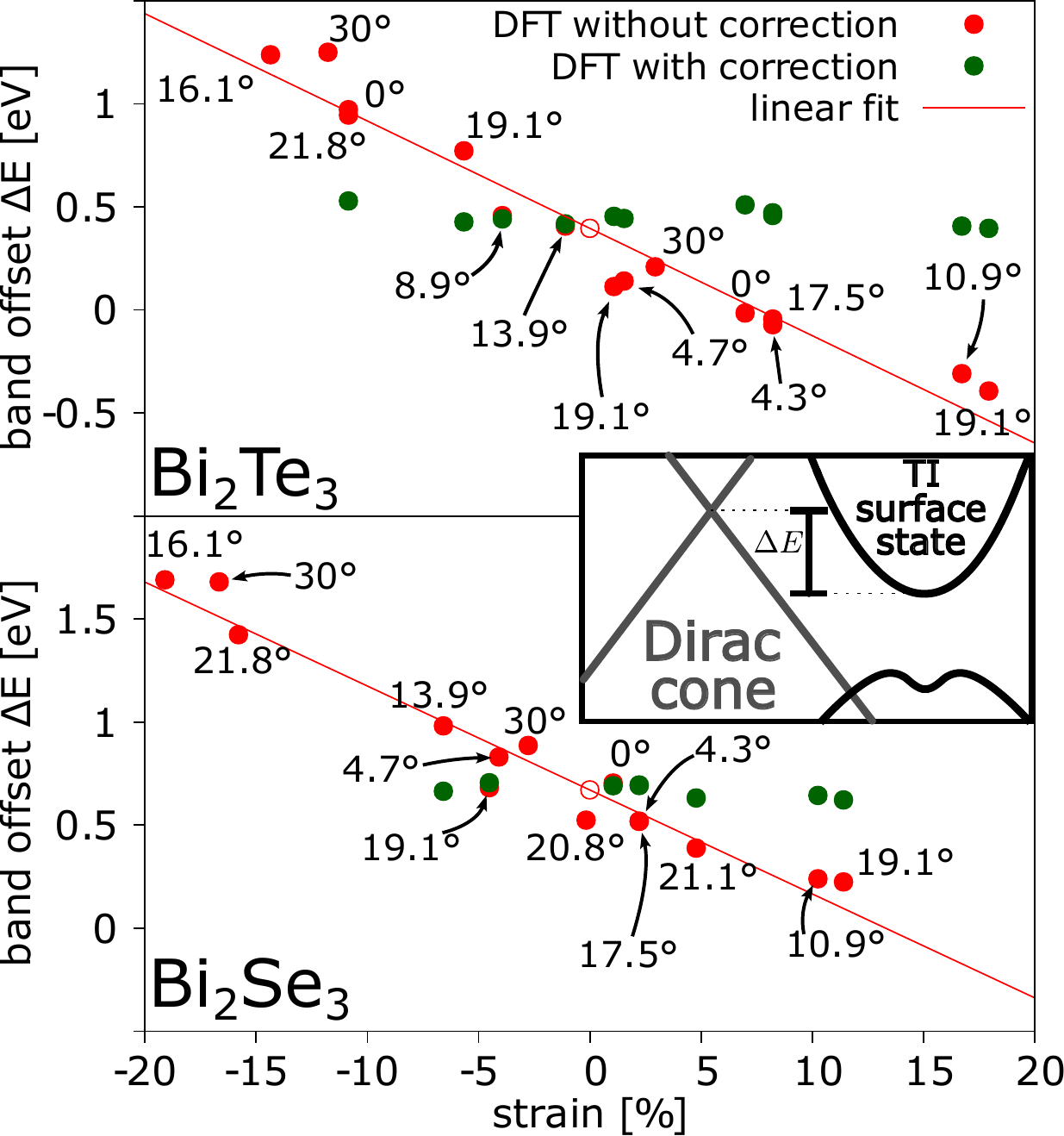} 
     \caption{Correcting for the strain induced band offset changes. For all the investigated supercells of graphene with 1QL of Bi$_2$Te$_3$  and Bi$_2$Se$_3$  we plot the band offsets $\Delta E$ of the Dirac cone with respect to the remnant of the TI surface state  (see inset) against the strain on graphene $\epsilon$; $\epsilon >0$ indicates tensile strain while $\epsilon<0$  indicates compressive strain. Each of the data points (red solid circles) is annotated with the twist angle of the corresponding supercell.  From the linear fit (red line) we extract the (apparent) zero-strain band offsets (empty red circles). The green circles show the band offsets after the correction by the transverse electric field employed to compensate the influence of strain. The inset shows schematically how the band offset $\Delta E$ is measured. %Strains above 10$\%$ and below -10$\%$ and negative band offsets $\Delta E < 0$ were not included in the linear fit.
     %{\color{red} Labels a and b are not in the figures. So it is best to remove them in the caption. Also, the inset should be described in the caption. check }
     }\label{Fig:strainVSoffset}
    \end{figure}

In Ref.~\citep{Naimer2021:paper1} we investigated graphene/TMDC heterostructures and reported a linear connection between the strain $\epsilon$ (enforced on the graphene) and the band offset between the graphene and the transition-metal dichalcogenide (TMDC) band structure. For the graphene/TI heterostructures we also observe a linear relation, see Fig.~\ref{Fig:strainVSoffset}, allowing us to estimate the apparent zero-strain band offset for both cases: $\Delta E=396$ meV for Bi$_2$Te$_3$ and $\Delta E=671$ meV for Bi$_2$Se$_3$. We then apply a transverse electric field to each supercell to reduce the band offset to the zero-strain one. 

As a reference point for the TI energies we use the TI surface state (or for the 1QL case: the remnants of the surface state) at $\Gamma$. Unlike thin TMDC monolayers, the TI multilayers we use are rather thick, having a thickness of $\approx$7 \AA~per QL. This makes them more vulnerable for unwanted side effects of the electric field to the band structure. A prominent example is the splitting of the TI surface state~\citep{Zollner2021:PSSB} into a state living at the lower TI surface (close to the graphene monolayer) and one living at the upper TI surface (further away from the graphene monolayer). However,  we expect the  consequences for the proximity SOC to be rather minimal, since the proximity SOC is induced mainly by the atomic orbitals close to the graphene monolayer in real space. %Therefore the band offsets are always measured with respect to the surface state living at the lower TI surface (close to the graphene monolayer).

The computational methodology for obtaining DFT electronic band structures of the graphene/TI supercells is detailed in App.~\ref{App:comp}.

%------------------------------------------------------------------------
%------------------------------------------------------------------------

%------------------------------------------------------------
\section{Qualitative picture of Interlayer interaction in k space}
\label{Sec:Umklapp}
\begin{figure}[htb]
     \includegraphics[width=.99\linewidth]{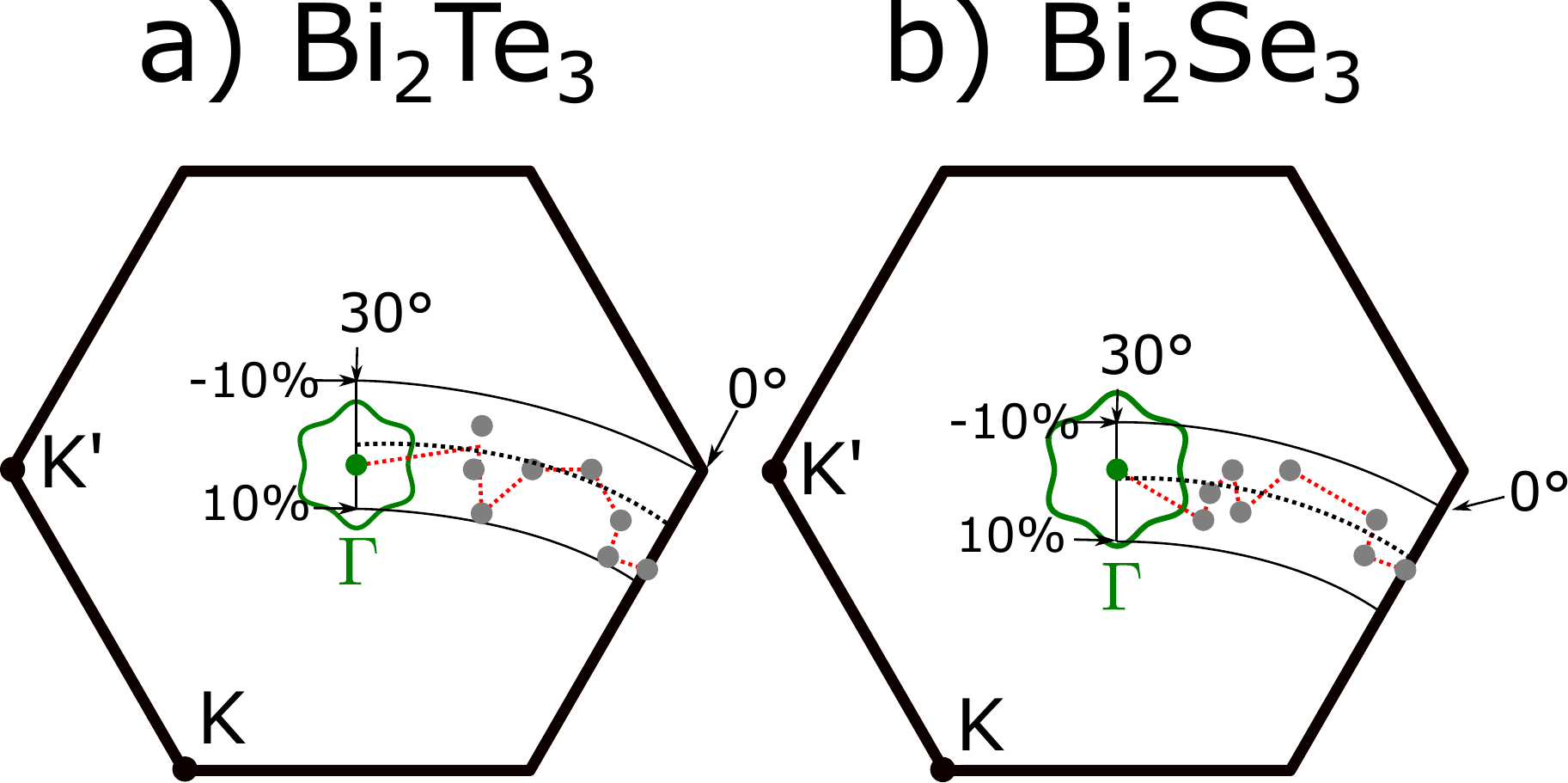} 
     \caption{
     Two dimensional first BZ of (a) Bi$_2$Te$_3$ and (b) Bi$_2$Se$_3$. The green dot marks the $\Gamma$-point, with the green line indicating the cross section of the TI surface state (for the 1 QL case) at the height of the electric field corrected Dirac cone energy.
     % {\color{red} Graphene Dirac point of corrected supercell? Yes,check}.
     For each calculated supercell the $\mathbf{k}$-points to which the Dirac cone couples are drawn (after symmetry reduction). Additionally, the zero-strain line is drawn (dotted line), formed by all points of hypothetical supercells with $\epsilon =0\%$ and $0\degree\leq\Theta\leq\ 30\degree$. The outlined sector indicates all relevant $\mathbf{k}$-points for a range of $ -10 \% \leq \epsilon \leq 10 \% $ and $0\degree\leq\Theta\leq\ 30\degree$. The red line connects all DFT data points with $\epsilon<10 \% $ and represents the sequence of points in Fig.~\ref{Fig:params}.
     }\label{Fig:hexagons}
    \end{figure}

%------------------------------------------------------------
Ref.~\citep{Koshino2015:TwistTBBasic} details by generalized Umklapp processes how in twisted heterostructures only certain $\mathbf{k}$-points of the two layers can interact with each other. In the graphene/TI heterostructures our focus is on the graphene low energy Dirac states. Therefore we are interested in the $\mathbf{k}$-points in the 2D primitive TI (Bi$_2$Se$_3$ or Bi$_2$Te$_3$) unit cell, with which the Dirac cone of graphene will primarily interact (and obtain its SOC from) in a graphene/TI heterostructure. The principle contribution comes from three $\mathbf{k}$-points, which are equivalent due to symmetry. % There are other $\mathbf{k}$-points, where a higher order interaction takes place. However, the SOC obtained from second order $\mathbf{k}$-points is already diminished by roughly two orders of magnitude and therefore negligible~\cite{Koshino2015:TwistTBBasic}. {\color{red} Give reference supporting this statement. check} 
The location of these $\mathbf{k}$-points depends both on the twist angle between the two materials and the ratio of their lattice constants, so in this case $a_{TI}/a_{Gr}$. Since the strain we apply to the graphene in order to construct a commensurate heterostructure changes $a_{TI}/a_{Gr}$, we can identify strain and twist angle as the two relevant factors for our calculations. %Assuming realistic strains in experiments, the effect of the strain becomes negligible.
Fig.~\ref{Fig:hexagons} shows for both Bi$_2$Se$_3$ and Bi$_2$Te$_3$, where those $\mathbf{k}$-points lie for the supercells listed in Tab.~\ref{Tab:Structures}.
%{\color{red} This paragraph is rather difficult to read. It is not clear what is meant by 1st and 2nd order k points, for example. Either drop that or explain more. It is also not clear why the effect of strain becomes negligible for $< 3\%$. Where does this number come from? check. I removed most of this argument}

As is also clear from Fig.~\ref{Fig:hexagons}, for the $30\degree$~supercell the graphene Dirac cone interacts exactly with the $\Gamma$-point of the TI. Since this is the reciprocal lattice momentum at which the surface states of the TI reside, this particular twist angle is expected to be special. Indeed, Ref.~\citep{Song2018:TIGRheteroDFT} reports SOC of the Kane-Mele type appearing in DFT calculations on a graphene/Bi$_2$Se$_3$ heterostructure with a 30\degree~twist angle. 
We dedicate Subsec.~\ref{Subsec:ResII} to discussing this special case.

%------------------------------------------------------------------------
%------------------------------------------------------------------------

%------------------------------------------------------------
\section{Effective Hamiltonian}
\label{Sec:effHam}
%------------------------------------------------------------
In order to find the twist-angle dependence of the proximity induced SOC in graphene's Dirac bands due to the coupling with the TIs, we fit the DFT band structures at the Dirac points to a model
Hamiltonian~\citep{Gmitra2015:TMDCgraphene1}. The Hamiltonian $H$ comprises the orbital part $H_{\text{orb}}$ and the spin-orbit part $H_{\text{so}}$. The latter is composed of the 
intrinsic spin-orbit coupling $H_{\text{so,I}}$ and the Rashba coupling $H_{\text{so,R}}$:
\begin{equation}
H(\mathbf{k})=H_{\text{orb}}(\mathbf{k})+H_{\text{so}}=H_{\text{orb}}(\mathbf{k})+
H_{\text{so,I}}+H_{\text{so,R}}.
\label{Eq:Ham}
\end{equation}
The orbital part describes the dispersion of the graphene Dirac cone linearized around the $K$/$K'$-point; accordingly, $\mathbf{k}$ is the electron wave vector measured from $K$/$K'$. It also includes a staggered potential $\Delta$, caused by the substrate's asymmetrical influence on the graphene A- and B-sublattice:
\begin{equation}
H_{\text{orb}}(\mathbf{k})=\hbar v_F (\kappa\sigma_x k_x+\sigma_y k_y)+\Delta \sigma_z.
\end{equation}
Here, $v_F$ is the Fermi velocity of the Dirac electrons and $\sigma$ are the Pauli matrices operating on the sublattice (A/B) space. The parameter $\kappa= 1$ for $K$ and 
$\kappa=-1$ for $K'$.

The intrinsic spin-orbit Hamiltonian
\begin{equation}
H_{\text{so,I}}=\Big[\lambda_{\text{KM}}\sigma_z+\lambda_{\text{VZ}}\sigma_0\Big]\kappa s_z,
\label{Eq:HamKMVZ}
\end{equation}
 and the Rashba spin-orbit Hamiltonian
 \begin{equation}
H_{\text{so,R}}= -\lambda_{\text{R}} \exp(-i\Phi \frac{s_z}{2})
\Big[\kappa\sigma_x s_y-\sigma_y s_x\Big]\exp(i\Phi \frac{s_z}{2}),
\label{Eq:HamR}
\end{equation}
both additionally act on the spin space, which is described by the spin Pauli matrices $s_x,s_y$ and $s_z$; $\lambda_{\text{VZ}}$ and $\lambda_{\text{KM}}$ are the valley-Zeeman~\citep{Gmitra2015:TMDCgraphene1, Wang2015:NC} SOC (sublattice-odd) and the Kane-Mele~\citep{Kane2005:PRL} SOC (sublattice-even) respectively. The Rashba SOC term is defined by a magnitude $|\lambda_{\text{R}}|$ and a phase angle $\Phi$. The latter is present in $C_3$ symmetric structures~\citep{Li2019:TwistTB1, David2019:TwistTB2, Naimer2021:paper1} and rotates the spin texture about the $z$-axis, adding a radial component to the Rashba field.

We choose to limit the Rashba parameter to positive values $\lambda_{\text{R}} >0$. A sign change of $\lambda_{\text{R}}$ then corresponds to an additional phase shift of $\Phi$ by a half rotation, i.e. $\Phi\rightarrow\Phi+180\degree$. To make this clear we always write $|\lambda_{\text{R}}|$.

We only construct heterostructure with angles between 0\degree~and 30\degree. The parameters for all other twist angles can be obtained by the following symmetry rules:

Twisting clockwise or counterclockwise from 0\degree~influences only the Rashba phase angle: 
\begin{align}
\lambda_{\text{VZ}}(-\Theta)&=\lambda_{\text{VZ}}(\Theta) \\
|\lambda_{\text{R}}(-\Theta)|&=|\lambda_{\text{R}}(\Theta)|\\
\Phi(-\Theta)&=-\Phi(\Theta)\\
\Delta(-\Theta)&=\Delta(\Theta).
\end{align}

Additionally a twist by 60\degree~corresponds to switching the sublattices of graphene and therefore changes the sign of the sublattice-sensitive parameters:
\begin{align}
\lambda_{\text{VZ}}(\Theta+60\degree)&=-\lambda_{\text{VZ}}(\Theta)\\
|\lambda_{\text{R}}(\Theta+60\degree)|&=|\lambda_{\text{R}}(\Theta)|\\
\Phi(\Theta+60\degree)&=\Phi(\Theta)\\
\Delta(\Theta+60\degree)&=-\Delta(\Theta).
\end{align}

%{\color{red} To make the paper more self-contain we should reproduce the rules here. It is a long paper anyway.}

%------------------------------------------------------------
\section{Results}
\label{Sec:Res}
\begin{figure*}[htb]
     \includegraphics[width=.99\linewidth]{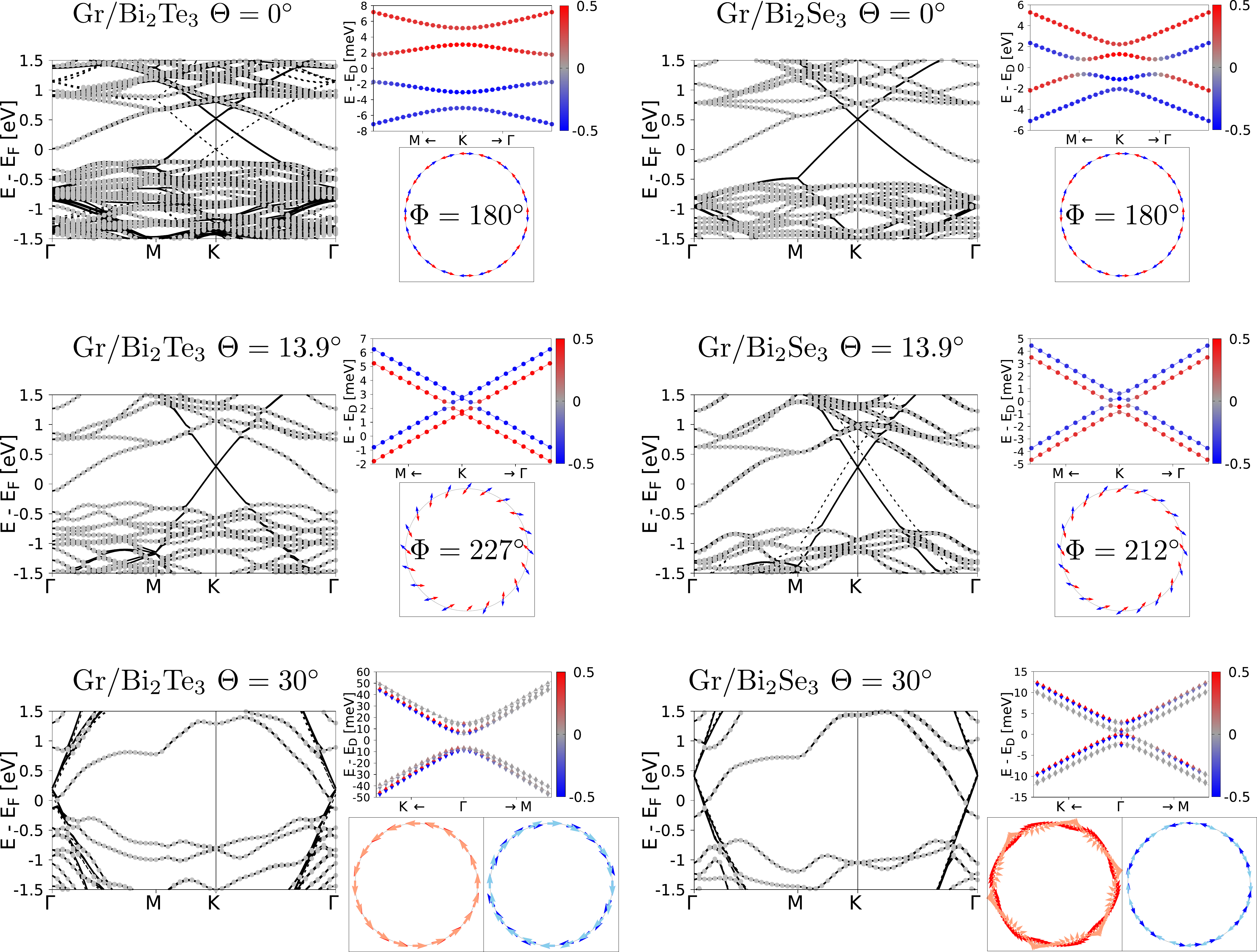} 
     \caption{Calculated band structures of graphene/TI heterostructures using Bi$_2$Te$_3$ (first column),  Bi$_2$Se$_3$ (second column), and three different selected twisting angles $\Theta=0\degree,13.9\degree,30\degree$. The grey circles indicate states originating from TI orbitals, while the solid (dashed) lines come from graphene states with (without) the electric field corrections. For the low-strain cases (Gr/Bi$_2$Te$_3$ 13.9\degree, Gr/Bi$_2$Se$_3$ 0\degree~and both 30\degree~cases) %{\color{red} WHICH angles?.check.})
     the solid and dashed lines coincide, because no strong shifting electric field is needed. The origin of the energy scale is set to zero for the Fermi energy of the band structures without electric field correction. The remnant of the TI surface state can be seen at $\Gamma$. Because of band folding effects, the graphene Dirac cone resides at the $K$-point for all angles except $\Theta=30\degree$. In addition to the band structures along high symmetry lines, we show zooms to the (electric-field corrected) Dirac cone with color coded spin and the in-plane spin-structure along a circular path around the Dirac cone at $\approx 55$~meV. For the zooms, dots show DFT data while solid lines represent the fits from the model Hamiltonian (Eq.~\eqref{Eq:Ham}). In the circular-path plots, red (blue) arrows indicate in-plane spin in the energetically lower (higher) valence band. For $\Theta=0\degree$ the colors coincide with spin-z, but (due to the sign change in $\lambda_{VZ}$) they do not for $\Theta=13.9\degree$. Since for $\Theta=30\degree$ the band structure comprises 8 bands, we show two plots with red and orange (blue and light-blue) arrows indicating the energetically lower (higher) pair of valence bands. Conduction bands show the same structure.  The Rashba phase angles $\Phi$ are extracted for all, but the $\Theta=30\degree$ case. Note that for the zoomed band structures of the $\Theta=30\degree$ cases the $k$ window is enlarged by a factor of 3 (Bi$_2$Se$_3$) or 15 (Bi$_2$Te$_3$).% {\color{red} The labels model fit and DFT should have bigger fonts. You can also remove the labels, as the symbols are described in the caption.check.}
    }\label{Fig:GlobalBS}
    \end{figure*}

We calculate the electronic band structures for all supercells listed in boldface in Tab.~\ref{Tab:Structures} by means of DFT for the 1QL case. Fig.~\ref{Fig:GlobalBS} shows a few representative band structures. In addition it shows zooms to the graphene Dirac cone and the in-plane spin structure around it. It is clearly visible that the proximity SOC and therefore the structure of the Dirac cone is different for different angles. Fitting the Dirac cones to the model Hamiltonian, Eq.~\eqref{Eq:Ham}, results in effective model parameters (see Tab.~\ref{Tab:param}) which can be used to compare different twist angles. 
 In the following, we discuss these results separately for twist angles below 20$\degree$ and for the twist angle at 30$\degree$, as the two cases have distinct features. 
%------------------------------------------------------------------------
\subsection{Results I: $0\degree\leq\Theta\lessapprox 20\degree$ }
\label{Subsec:ResI}
%------------------------------------------------------------

%------------------------------------------------------------------------
For twist angles $0\degree\leq\Theta\lessapprox 20\degree$ the graphene Dirac point acquires its SOC mainly from parts of the TI first BZ which are away from the $\Gamma$-point. The corresponding Bloch states are not what would 
form TI surface states,  as those appear at $\Gamma$. Hence, the overall SOC is weaker than for the $\Theta=30\degree$ case (see next subsection).  Additionally, the SOC has similar functional form to that in graphene/TMDC heterostructures~\citep{Naimer2021:paper1}: both staggered potential $\Delta$ and $\lambda_\text{KM}$ are negligibly small, while $\lambda_\text{VZ}$ and $|\lambda_R|$ dominate. This is in agreement with earlier calculations~\citep{Zollner2021:PSSB,   Song2018:TIGRheteroDFT}. Also, the Rashba phase angle vanishes ($\Phi=180\degree$) for $\Theta=0\degree$ due to symmetry.

Fig.~\ref{Fig:params} depicts the twist-angle dependence of the extracted SOC parameters $\lambda_{VZ}$, $|\lambda_{R}|$, and $\Phi$. The qualitative structure is the same for both materials Bi$_2$Se$_3$ and Bi$_2$Te$_3$.
It exhibits a special feature of the graphene/TI heterostructures, namely, the sign change of $\lambda_{VZ}$ at about $\Theta=10\degree$. Since increasing the TI thickness from 1QL to 3QL leaves the sign of 
$\lambda_{VZ}$ unaffected (see App.~\ref{App:QL}), it is reasonable to conjecture, taking into account the short range of the proximity effect in van der Waals heterostructures, that the valley-Zeeman SOC changes sign also for graphene on bulk TI.

Since heterostructures with large strain ($\epsilon<10\%)$ couple to very different parts of the TI Brillouin zone, they deviate strongly from the zero-strain path in Fig.~\ref{Fig:hexagons} and are therefore depicted as transparent points in Fig.~\ref{Fig:params}. Nevertheless, we can infer from these calculation that there can be a sign change not only by sweeping the twist angle $\Theta$, but also by sweeping the strain $\epsilon$. E.g., a graphene/Bi$_2$Te$_3$ heterostructure with a fixed twist angle of 19.1\degree~changes the sign of   $\lambda_\text{VZ}$, when going from moderate strains ($\epsilon_1$=-5.66\%,  $\epsilon_2$=1.08\%) to a very large positive strain ($\epsilon_3$=17.93\%). We estimate this sign change to happen at 8\% to 10\% for twist angles $\Theta>10\degree$. Although such heterostructures are not directly realizable in an experiment (due to the large strain), it is a relevant side note for heterostructures using similar TIs or alloys of TIs, which have slightly different lattice constants. The sign changes are visualized in the second line of Fig.~\ref{Fig:params}.

Remarkably, there are two irregularities that appear. Firstly, there is a rather abrupt sign change in $\lambda_\text{VZ}$ between the two last data points ($\Theta=20.8\degree$ and $\Theta=21.1\degree$) of the Bi$_2$Se$_3$ heterostructures in Fig.~\ref{Fig:params}(b). This could be due to the close vicinity of the Dirac cone of the $\Theta=21.1\degree$ heterostructure to the TI surface state: Despite the rather small difference in twist angle, due to its different strain its Dirac cone couples to a point closer to the $\Gamma - M$ line rather than the $\Gamma - K$ line. Since in this direction in $k$ space the slope of the TI surface state is less steep, it is closer to the Dirac cone in energy. And apparently this influence manifests in the sign change of $\lambda_{VZ}$. Secondly, The magnitude $|\lambda_R|$ of the Rashba SOC is generally smaller than the magnitude of $\lambda_{VZ}$. It seems to almost monotonically decrease for $0\degree\leq\Theta\lessapprox 20\degree$. However, when looking again at the $\Theta=21.1\degree$ Bi$_2$Se$_3$ data point, we see a strong increase of $|\lambda_R|$, again likely related to the Dirac cone's vicinity to the remnants of the surface state.

%------------------------------------------------------------------------
    \begin{figure}[htb]
     \includegraphics[width=.99\linewidth]{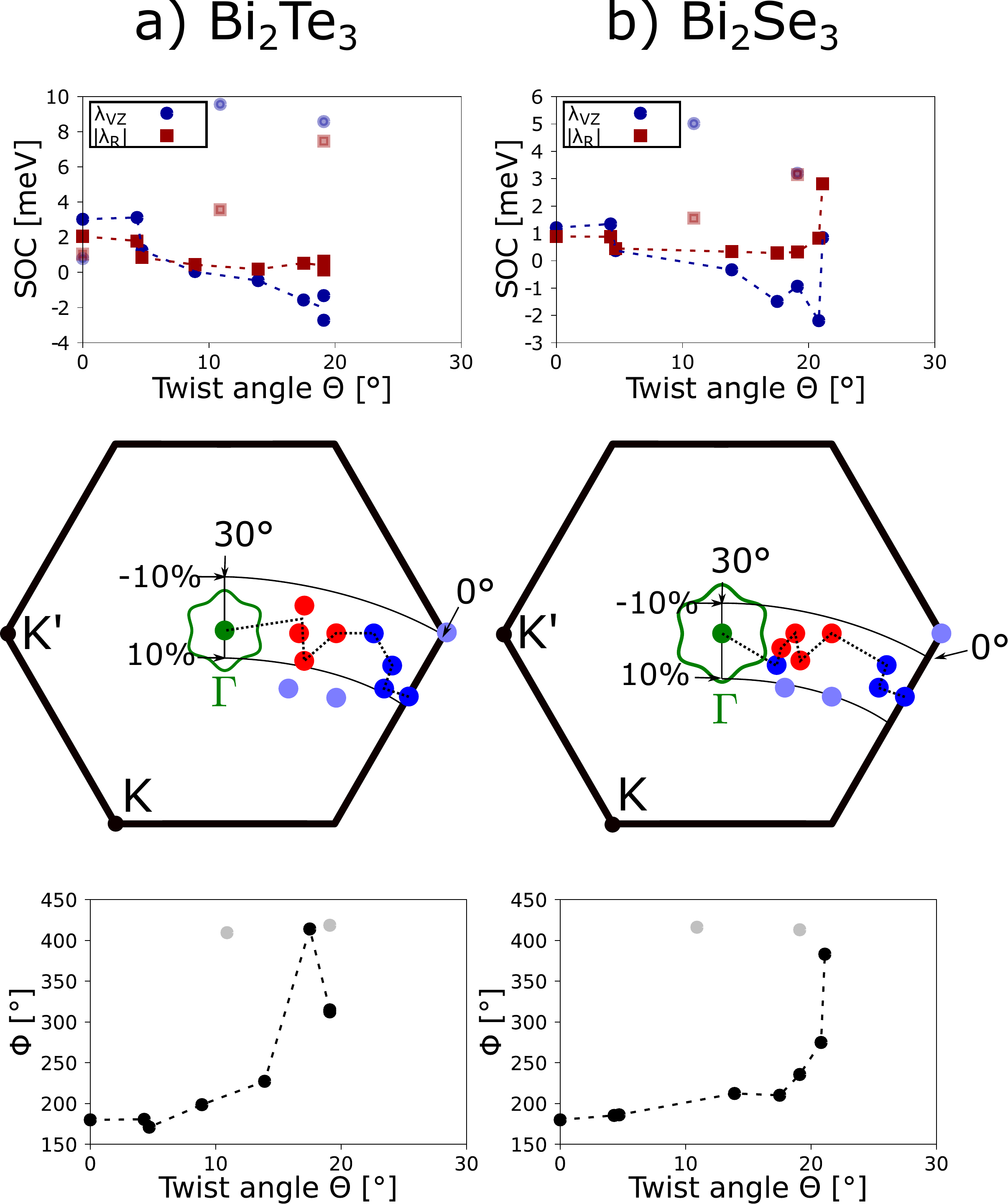} 
     \caption{
    Calculated SOC parameters of the graphene Dirac cone proximitized by (a) Bi$_2$Te$_3$ and (b) Bi$_2$Se$_3$. The effect of strain is corrected by transverse electric field, as described in Sec.~\ref{Sec:methods}. The first line of plots shows Rashba SOC $|\lambda_{\text{R}}|$ (red squares) and valley-Zeeman SOC $\lambda_{\text{VZ}}$ (blue circles) as a function of the twist angle $\Theta$. Data points from supercells with excessive built-in strain of $\epsilon>10\%$ are de-emphasized (transparent). The dotted line is a guide for the eyes.
    The second line of plots shows the TI Brillouin zone and filled circles indicating regions in which the Dirac cone couples to the TI Bloch states, as in Fig.~\ref{Fig:hexagons}. Blue symbols correspond to positive valley-Zeeman SOC $\lambda_{\text{VZ}}>0$, while red symbols correspond to negative valley Zeeman SOC $\lambda_{\text{VZ}}<0$. The green circle represents the $\Gamma$-point as well as the 30\degree~supercell connecting to it. The dotted line connects the relevant points with ascending twist angle in the same order as the guide to the eyes in the first line of plots. The third line of plots shows the twist-angle dependence of the Rashba phase angle $\Phi$. Again, data points with $\epsilon>10\%$ are de-emphasized (grey).
     \label{Fig:params}
     }\label{Fig:params}
    \end{figure}
%------------------------------------------------------------------------

The Rashba phase angle $\Phi$ is essential for collinear charge-to-spin conversion~\citep{Zhao2022:expradialrashba,Ingla_Aynes2022:expradialrashba2,Lee22:PRB:radialrashba3,Peterfalvi2022:radialrashba,Veneri22:PRB:radialrashba2}, %{\color{red} Something wrong with citations.check}
since for $\Phi=0\degree$ and  $\Phi=180\degree$ collinear charge-to-spin conversion is forbidden. The twist-angle dependence of $\Phi$ in graphene-based heterostructures is not well explored apart from the symmetry-dictated fact that $\Theta=0\degree$ and $\Theta=30\degree$ both entail $\Phi=0\degree$ or $\Phi=180\degree$~\citep{Li2019:TwistTB1,Peterfalvi2022:radialrashba,Naimer2021:paper1}.

For intermediate twist angles $0\degree<\Theta<30\degree$, there can be either no sign change $\Phi(\Theta=0\degree)=\Phi(\Theta=30\degree)$ or a sign change $\Phi(\Theta=0\degree)=\Phi(\Theta=30\degree)+180\degree$. The latter is especially interesting, since it implies the existence of a twist angle  $0\degree<\Theta<30\degree$ for which $\Phi=90\degree$ or $\Phi=270\degree$ and therefore a purely radial Rashba spin structure and purely collinear charge-to-spin conversion. Ref.~\citep{Peterfalvi2022:radialrashba} predicts  such a sign change to happen for certain graphene/TMDC heterostructures based on a tight binding model, although DFT results~\cite{Naimer2021:paper1,Lee22:PRB:radialrashba3} are at odds with that prediction. %{\color{red} Although DFT results (ref) and other TB calculations (ref) are at odds with that prediction. Check. However I am not aware of the TB calculations which are at odds with it. Maybe you are confusing it with the VZ sign change...}
Our results seem to strongly indicate that such a sign change could occur for graphene/TI heterostructures with the twist angle corresponding to a purely radial spin structure being $\Theta\approx18\degree$ (see Fig.~\ref{Fig:params}).
%{\color{red} The 30 degree case should be discussed in the next section. However, the description could be made more clear.check shifted to next subsection }

%---------------------------------------0 0 0 0 0 0 0 0 0 0 0 0 0 0 0 0 0 0 0 0 0 0 0 0 0 0 00 0 0 0 0 0 0 0 0 0 0---------------------

\subsection{Results II: $\Theta=30\degree$}
\label{Subsec:ResII}

By combining a $\sqrt{3}\times\sqrt{3}$ graphene supercell ($n=m=1$) and a 1$\times$ 1 TI supercell ($n'=0,m'=1$) one can create a heterostructure with a twist angle of $\Theta=30\degree$. Even though other heterostructures with such a twist angle can be constructed (e.g. $n=7,m=0,n'=m'=2$), the former is unique in a few ways:
\begin{enumerate}
    \item It is a notably small supercell. For larger heterostructures the shifting degree of freedom is mostly irrelevant for the proximity SOC, because the graphene will have many different local atomic registries, which will always result in some average proximity effect. Since this is impossible for such a small supercell, the shifting degree of freedom will strongly affect the low energy Dirac cone band structure (see App.~\ref{App:shift}).
    \item The $K$- and $K'$-points of the primitive graphene first Brillouin zone are folding back to the $\Gamma$-point of the supercell's  first Brillouin zone, creating an 8 band Dirac cone.
    \item Not only will the Dirac cones fold back to the $\Gamma$-point, but, more significantly, the point of the TI BZ, with which the graphene Dirac cone will interact by the theory of generalized Umklapp processes (see Sec.~\ref{Sec:Umklapp} and Ref.~\citep{Koshino2015:TwistTBBasic}) is exactly the $\Gamma$-point, where  the TI surface state resides.
\end{enumerate}
According to the first point, we observe very different low energy spectra for the four different shifting configurations (see Fig.~\ref{Fig:shift}). For all but the `Hollow' configuration we see eight distinct bands, degenerate only at the $\Gamma$-point, where Cramer's rule strictly demands it. The 'Hollow' configuration is the energetically most favourable one and therefore the one we will focus on (all plots in the main manuscript regarding the $\Theta=30^\circ$ supercell represent this 'Hollow' case). It entails band structures consisting of four band pairs with energy splittings within such a band pair being on the $\mu$eV range. These small splittings lead to a certain spin structure (see App.~\ref{App:finestruc}), that has also been found and discussed in Ref.~\citep{Song2018:TIGRheteroDFT}. % In App.~\ref{App:spinstruc} we describe this spin structure using a full tight binding Hamiltonian akin to the one of Song et al.~\citep{Song2018:TIGRheteroDFT}, which importantly includes a handcrafted Rashba SOC based on in-plane electric fields.
However, since the splitting is very small and the resulting spin structure is very elusive, we only use the simple Hamiltonian (Eq.~\eqref{Eq:Ham}) described in Sec.~\ref{Sec:effHam} for the fittings in the main manuscript (we reduce the eight bands to four by ignoring one band of each almost-degenerate band pair). This on the one hand allows for better comparability with the parameters of the other twist angles, but on the other hand fails to describe the spin structure. In App.~\ref{App:Aaronfitting} we fit the Dirac cones with an alternative fitting Hamiltonian which is akin to the one used in Ref.~\cite{Song2018:TIGRheteroDFT} and find that the fitting parameters are very similar.

Due to the aforementioned backfolding, we cannot include spin-z expectation values in the fitting procedure. However, pseudospin (sublattice imbalance) is used in addition to the energies to unambiguously determine the correct parameters. For the `Hollow' case pseudospin is always zero, therefore demanding $\Delta=\lambda_{VZ}=0$ (this can alternatively be deduced from symmetry). For the `Top' case the pseudospins of the eight bands form a complicated structure, which can be roughly reproduced using a full tight binding model Hamiltonian including certain onsite potentials (see App.~\ref{App:shift}). The energies and pseudospins of the `Bridge' case could not be sufficiently reproduced by either Hamiltonian.

The in-plane spin structure of the graphene/Bi$_2$Se$_3$ 30\degree~case (see Fig.~\ref{Fig:GlobalBS}, right column, last line) can clearly not be described with the Hamiltonian in Eq.~\eqref{Eq:Ham}.
For the graphene/Bi$_2$Te$_3$ 30\degree~case (see Fig.~\ref{Fig:GlobalBS}, left column, last line) the band pairs seem to exhibit a typical tangential Rashba in-plane spin structure. However, the order in which clockwise (c) or counterclockwise (cc) spin structures appear (starting from the energetically lowest band pair) is alternating, i.e. cc-c-cc-c. For the usual Rashba case it is cc-c-c-cc. Therefore, the in-plane spin structures also cannot be described by Eq.~\eqref{Eq:Ham} and we cannot estimate a Rashba phase angle from such calculations. This means there appears to be physics in the commensurate system, which the simple model Hamiltonian (Eq.~\eqref{Eq:Ham}) is not able to capture. Assuming that this physics stems from the specific atomic registry dependence, an unstrained incommensurate structure should still be well described by the simple model Hamiltonian (Eq.~\eqref{Eq:Ham}). % {\color{red}ref to equation for the Hamiltonian. Why capital M?.check.}. 
Based on the results from last line of plots of Fig.~\ref{Fig:params}, we presume that the phase angle $\Phi$ of such an incommensurate structure will be shifted by $180\degree$ with respect to the one at $\Theta=0\degree$.  Experimentally, a commensurate structure could be the result of chemical vapor deposition (CVD) or molecular beam epitaxy (MBE) fabrication, while the incommensurate structure might be obtained by a exfoliation method.

For $0\degree\leq\Theta\lessapprox 20\degree$ the Dirac cone lies (locally) within a band gap. As already mentioned for $\Theta=30\degree$ the Dirac cone now lies directly on top of the TI surface state at the $\Gamma$-point. Therefore, both the thickness of the TI (determining the concrete form of the surface state) and an applied transversal electric field (determining the relative position of the Dirac cone with respect to the surface state) can be expected to strongly influence the proximity SOC induced in the graphene. In the following, we will focus on the `Hollow' shifting configuration, since it is the energetically most favourable. 

\begin{figure*}[htb]
     \includegraphics[width=.99\linewidth]{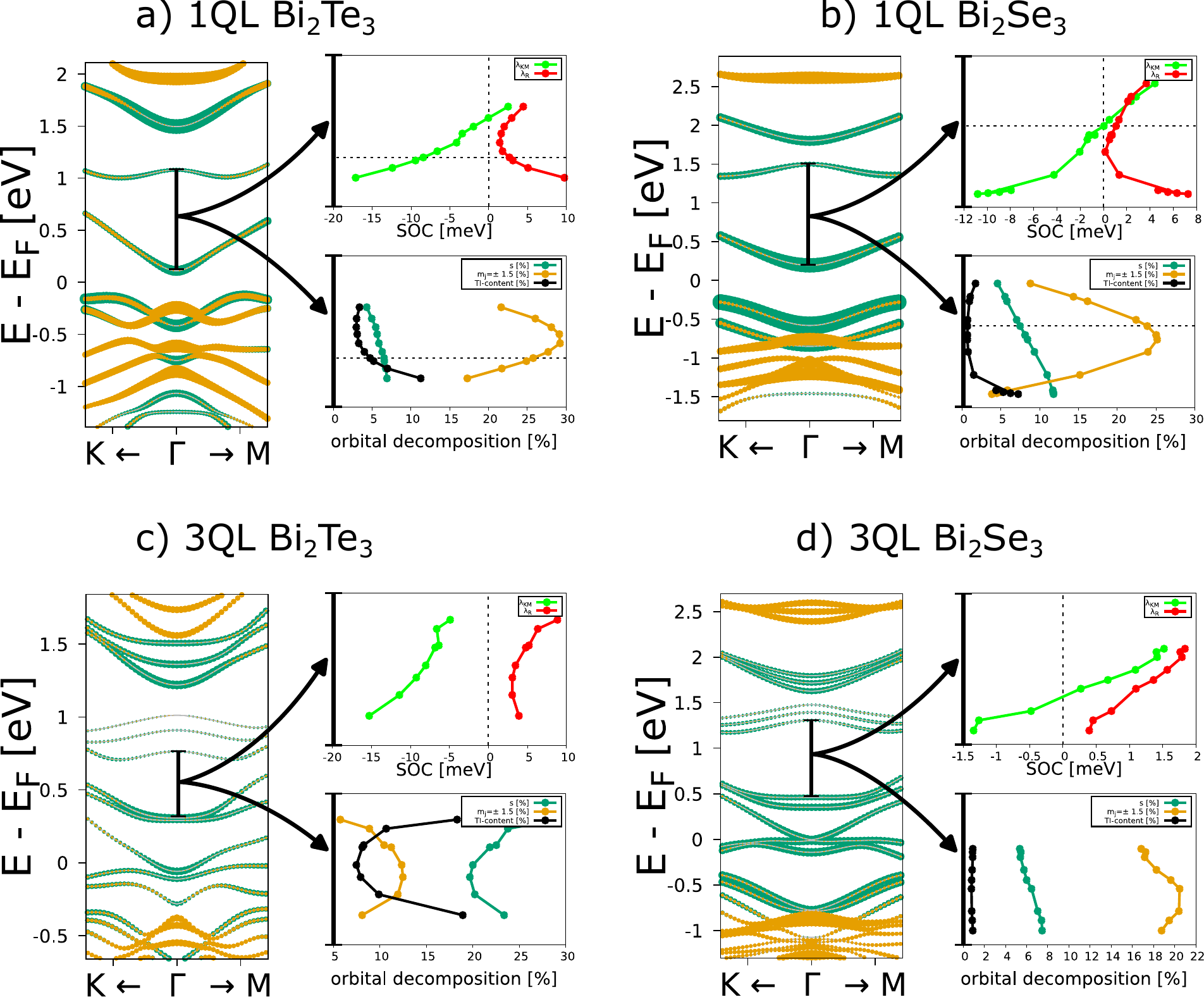} 
     \caption{Electric field dependence of fitting parameters for the 30\degree~supercell: We show the band structure of 1QL or 3QL of Bi$_2$Te$_3$ or Bi$_2$Se$_3$ around the $\Gamma$-point along high symmetry lines. The color code represents the projection of the state onto the TI s-orbitals (green) and onto the p-orbitals  with $m_j=\pm  3 / 2$ (yellow) respectively. The black range shows the energy range within which we shift the graphene Dirac cone in the respective Gr/TI heterostructures using an electric field. We show the development of the SOC parameters and the orbital decomposition of the Dirac cone states (averaged over a $k$ range of $0.04\frac{1}{\textbf{\AA} }$ for Bi$_2$Te$_3$ and $0.004\frac{1}{\textbf{\AA} }$ for Bi$_2$Se$_3$ along the high symmetry path around $K$) during this shifting. For a) and b) the horizontal dotted lines indicate the Dirac cone position of an electric field corrected band structure (zero-strain band offset), therefore corresponding to the zooms in the last line of plots in Fig.~\ref{Fig:GlobalBS}. In the lower plot ('orbital composition') the black curve shows the general content of TI-orbitals in the Dirac cone. The dark-green curve shows, how much of those states are TI s-orbitals. The yellow curve shows, how much of the contributing TI p-orbitals have quantum number $m_j=\pm 3 / 2$.}
     \label{Fig:ResII}
    \end{figure*}

%{\color{red} Perhaps we could place Fig. 5. after Fig. 2.? There is essential information there and the figure is not referenced in the text, it looks like, but only in the Appendix. I swapped it with fig.4 }
In Fig.~\ref{Fig:ResII} we show the electric field dependence of the parameters for the 1QL case and the 3QL case. The Dirac cone is shifted through the range marked in black within the TI band structure using an electric field. With $\Delta=\lambda_{VZ}=0$ as stated before, the two remaining parameters $\lambda_{KM}$ and $|\lambda_{R}|$ are both in the meV range. When the Dirac cone comes close to a TI band, the SOC becomes larger with both parameters reaching up to 20 meV in magnitude. Parameter $\lambda_{KM}$ can remarkably change sign whilst the Dirac cone moves from one band to another. Additionally, we depict the orbital decomposition of the Dirac cone. The black curve, showing the general TI-content and therefore the strength of the proximity effect will unsurprisingly increase when nearing one of the TI bands. We find TI p-orbitals to be contributing the most to the proximity effects ($\approx 90\%$), while s-orbitals contribute to a less, but still significant, amount ($\approx 10\%$). The dark green curve shows that this distribution stays roughly constant, when shifting the Dirac cone, with a tendency for higher s-orbital content near the surface state. The d-orbitals contribution is negligibly small. Distinguishing even further, we investigated the percentage of p-orbitals with $m_j=\pm 3 / 2$ and those with $m_j=\pm  1 / 2$. This distinction is essentially the equivalent of distinguishing between p$_{x/y}$- and p$_z$-orbitals in a spinless case, only now that SOC is present $m_l$ is not a good quantum number and must be replaced with $m_j$. We find that the $m_j=\pm  3 / 2$ orbitals make up roughly 20\% of the contributing p-orbitals, which is rather high, considering the Dirac cone p$_z$ orbitals overlap more with other p$_z$ orbitals and that the TI bands near the Dirac cone consist of hardly any states with $m_j=\pm  3 / 2$ . We conclude that an astonishingly large contribution to the proximity effect is coming from bands more than 2 eV away from the Dirac cone in the valence and conduction band, rather than from the surface state itself. Therefore, when the Dirac cone approaches any nearby band with low $m_j=\pm  3 / 2$ content, it will acquire a significant proximity effect from it and the relative contribution of the deep-lying states decreases. This corresponds to a  decline of the yellow curve indicating relative $m_j=\pm  3 / 2$ p-orbital content.

Another interesting feature of the band structure is the separation into two distinct kinds of band pairs, occurring in an alternating fashion. One of those kind of band pairs inherits its properties (mainly spin structure and orbital composition) from the nearby surface state, while the other one inherits its properties from the deep-lying states. This distinction is described in App.~\ref{App:finestruc} in detail.

%------------------------------------------------------------
%------------------------------------------------------------

%{\color{red} Make sure that all the Appendiced are referred and briefly summarized in the main text, not only in the introduction.}
\section{Summary}
\label{Sec:Sum}
We systematically investigated the proximity induced SOC in twisted graphene/Bi$_2$Se$_3$ and graphene/Bi$_2$Te$_3$ heterostructures. After determining an approximate zero-strain band offset, we correct the band offset of all structures accordingly and fit their energies and spin to an established SOC Hamiltonian to extract relevant SOC parameters. We separately consider supercells in a twist-angle range $0\degree\leq\Theta\lessapprox 20\degree$, which are barely affected by the TI surface state, and one special highly commensurate supercell at $\Theta=30\degree$, which is heavily influenced by the surface state. For the $0\degree\leq\Theta\lessapprox 20\degree$ supercells we extract the twist-angle dependence of the relevant types of SOC, which are valley-Zeeman ($\lambda_{VZ}$) and Rashba ($|\lambda_R|$, $\Phi$). Upon twisting, we witness a change of the valley-Zeeman sign at $\Theta\approx 10\degree$. Additionally we witness a sign change upon changing the strain at about +8\% to +10\% strain for twist angles $\Theta>10\degree$. We confirm that the Rashba phase angle $\Phi$ has a value of $\Phi=180\degree$ at $\Theta=0\degree$. For increasing twist angle the phase angle $\Phi$ also increases and crosses $\Phi=270\degree$ for $\Theta\approx 18\degree$, where a purely radial in-plane spin structure occurs. Assuming that this trend continues for twist angles $\Theta>20\degree$, it seems likely that (for incommensurate heterostructures) the phase angle will have a value of $\Phi=360\degree$ at a twist angle of $\Theta=30\degree$.

The case of the highly commensurate $\Theta=30\degree$ supercell was investigated more closely. We consider the specific shifting configuration 'Hollow', since it is energetically the most favourable. Since its symmetries do not allow for valley-Zeeman SOC, the significant SOC types are Rashba ($|\lambda_R|$) and Kane-Mele ($\lambda_{KM}$).
Due to backfolding effects, the Dirac cone coincides with the TI surface state in $k$ space. Using an electric field, we shift the Dirac cone within a local TI band gap, which results in a change of the sign of $\lambda_{KM}$. Furthermore, orbital decomposition considerations reveal that the Dirac cone consists of two distinct alternating types of bands, one of which obtains its proximity SOC almost exclusively from higher lying bands.

In addition, we show calculations indicating that the effect of lateral shifting is irrelevant for the cases within $0\degree\leq\Theta\lessapprox 20\degree$, while having a strong effect on the highly commensurate $\Theta=30\degree$ supercell.
%------------------------------------------------------------
 \begin{table*}[htb]
   \caption{Parameters extracted from the band structure calculations. For both Bi$_2$Te$_3$ and Bi$_2$Se$_3$ and for all angles (except if the band offset is too large and the Dirac cone is shifted into the TI bands), we list the band offset $\Delta E$  of the Dirac cone with respect to the TI surface band and the extracted model Hamiltonian (Eq.~\eqref{Eq:Ham}) parameters. The parameters are staggered potential $\Delta$, Kane-Mele SOC $\lambda_{KM}$, valley-Zeeman SOC $\lambda_{VZ}$, magnitude of the Rashba SOC $|\lambda_{\text{R}}|$ and Rashba angle $\Phi$. We denote the offsets and parameters after correction with the electric field with a bar, for example $\bar{\lambda}_{VZ}$. The electric field is defined as positive, if it points from the TI layer to the graphene layer.}
    \label{Tab:param} 
    \begin{ruledtabular}
    \begin{tabular}{cc|cccccc|c|cccccc}

$\Theta[\degree]$&$\epsilon$&$\Phi$&$\Delta$ & $\lambda_{KM}$ & $\lambda_{VZ}$ & $|\lambda_{\text{R}}|$& $\Delta E$& E-field &$\bar{\Phi}$&$\bar{\Delta} $&$\bar{\lambda}_{KM} $ & $\bar{\lambda}_{VZ} $ & $|\bar{\lambda}_R|$& $\bar{\Delta E}$ \\

&[\%]&[\degree]&[meV] & [meV] & [meV] & [meV]& [eV]& [$\text{V}/\text{nm}$]&[\degree]& [meV] &[meV]  & [meV] & [meV]& [eV] \\
\hline
Bi$_2$Te$_3$&&&&&&&&\\
\hline

0       &-10.85  &180           &2.325 & 0.698 & 1.256 &1.452      &0.972    &5.554    &180           &2.257 & 0.528 & 0.786 &1.037   &0.529              \\   
0       &6.98    &180           &0.002& -0.004 & 1.934 &0.748      &-0.015   &-3.399        &180           &0.037 & -0.022 & 3.026 &2.054 &0.510          \\
4.3     &8.22    &152           &0.000 & -0.006 & 2.038  &0.512     &-0.072   &-3.838    &181           &0.034  & -0.024 & 3.126 &1.791   &0.471      \\
4.7     &1.54    &146             &0.002 & -0.003 & 1.019 &0.557     &0.141   &-2.199      &171           &-1.936& -0.002 & 1.254 &0.852 &0.444          \\
8.9     &-3.93   &199           &0.418 & -0.002 & -0.319 &0.342        &0.458  &0.247     &199           &-0.001 & 0.008 & 0.050 &0.438  &0.442    \\
10.9    &16.72   &19           &-0.067 & 0.271 & 13.694 &7.140     &-0.309  &-6.171    &50           &0.109 & -0.082 & 9.562 &3.574    &0.407        \\       
13.9    &-1.1    &-130          &0.010 & 0.011 & -0.464 &0.178       &0.407   &-0.145   &-133          &0.010 & 0.006 & -0.467 &0.183    &0.418  \\
17.5    &8.22    &-10          &0.001 & 0.154 & 3.513 &2.897       &-0.044   &-3.624  &54           &0.014 & -0.028 & -1.568 &0.531     &0.458   \\
19.1    &1.08    &-42         &-0.001 & 0.024 & -0.814 &0.980       &0.114    &-2.406  &-45          &-0.001 & 0.005 & -1.320 &0.143      &0.114 \\
19.1    &17.93   &-           &- &- &- &-                            &-0.393  &-5.656   &59           &-0.034& -0.035 & 8.578 &7.467    &0.396           \\
19.1    &-5.66   &-          &- &- &- &-                            &0.772    &1.797   &-48          &0.007& 0.006 & -2.720 &0.636        &0.427          \\
30      &2.94        &-      &0.000 & -12.400 & 0.000 &5.089      & 0.255       & -0.771    &-&0.000& -8.413 & 0.000&2.703      &   0.324  \\

\hline
Bi$_2$Se$_3$ &&&&&&&&&\\
\hline

0       &-15.79  &180           &1.496 & 0.252 & 0.726 &0.816      &1.423    &-      &-           &- &- &- &-         &- \\               
0       &1.05    &180       &0.002 & -0.007 & 1.217 &0.901          &0.704    &0.074          &180           &0.003 & -0.007 & 1.204 &0.881    &0.692    \\           
4.3     &2.22    &-178          &-0.001& -0.006 & 1.195 &0.621     &0.518    &-1.36       &-175          &0.004 & -0.010 & 1.344 &0.879        &0.695  \\
4.7     &-4.09   &190           &0.001 & -0.005 & 0.381 &0.616      &0.831    &1.05    &-174           &-0.001& 0.000 & 0.369 &0.447          &0.592     \\
10.9    &10.25   &47           &-0.049 & 0.002 & 3.175 &1.024      &0.240    &-4.114     &56           &0.054 & -0.018 & 5.016 &1.554         &0.644 \\
13.9    &-6.58   &-155           &0.004 & 0.007 & -0.496 &0.623     &0.982    &2.216      &-148          &0.003 & 0.004 & -0.331 &0.330         &0.664    \\               
17.5    &2.22    &17           &0.080 & 0.027 & -0.552 &0.076       &0.519    &-1.354    &-150          &0.006 & -0.017 & -1.490 &0.281         &0.692        \\       
19.1    &11.4    &47           &0.035 & 0.006 & 2.304 &2.217       &0.225    &-3.624      &53           &0.003 & -0.035 & 3.197 &3.146         &0.622\\
19.1    &-4.52   &-120          &0.005 & 0.002 & -0.917 &0.299      &0.682    &-0.098     &-124          &0.000 & 0.000 & -0.935 &0.315         &0.706  \\               
20.8    &-0.17   &-41          &0.001 & -0.007 & -1.258 &0.468       &0.524    &-1.313    &-85          &-0.008 & -0.031 & -2.194 &0.819            &0.669      \\
21.1    &4.78    &23           &0.022& -0.028 & 0.600 &1.680       &0.388    &-2.36    &23           &-0.005& -0.074 & 0.844 &2.809          &0.632\\
30      &-2.77   &-   &0.000& 0.517 & 0.000       &1.340       & 0.754     & 0.617      &-      &0.000& 0.049 & 0.000 & 1.091      &0.700\\

\hline

    \end{tabular}
    \end{ruledtabular}
    \end{table*}

\acknowledgments This work was funded by the International Doctorate~Program Topological~Insulators of the Elite~Network of Bavaria, the Deutsche Forschungsgemeinschaft (DFG, German Research Foundation) SFB 1277 (Project-ID 314695032), SPP 2244 (project no. 443416183), and by the European Union Horizon 2020 Research and Innovation Program under contract number 881603 (Graphene Flagship), and Flagera project 2DSOTECH.

\appendix
%------------------------------------------------------------
\section{Computational Details}
\label{App:comp}
All electronic structure calculations
	are performed implementing density functional theory (DFT)~\citep{Hohenberg1964:PRB} 
	using {\tt Quantum ESPRESSO}~\citep{Giannozzi2009:JPCM}.
	Self-consistent calculations are carried out with a $k$ point sampling of $n_k\times n_k\times 1$. The number $n_k$ is listed in Table~\ref{Tab:ks} for all cases.
	We use charge density cutoffs $E_\rho=480$Ry and wave function kinetic cutoff $E_{\text{wfc}}=48$Ry ($E_{\text{wfc}}=58$Ry for Bi$_2$Se$_3$) for the fully relativistic pseudopotential
	with the projector augmented wave method~\citep{Kresse1999:PRB} with the 
	Perdew-Burke-Ernzerhof exchange correlation functional~\citep{Perdew1996:PRL}. Graphene's $d$-orbitals are not included in the calculations. We used Grimme D-2
    Van der Waals corrections~\citep{Grimme2006:JCC,Grimme2010:JCP,Barone2009:JCC}.
	
 The electric fields are implemented in the DFT calculations using a sawtooth potential in $z$-direction within the quasi 2D unit cell. The electric potential increases linearly in the area of the heterostructure and then falls rapidly in the vacuum. 
%-----------------------------------------------------------------
    \begin{table}[htb]
    \caption{Computational details: $k$ grid density (we used a $n_k\times n_k$ grid) for all calculations.}\label{Tab:ks}
    \begin{ruledtabular}
    \begin{tabular}{c|cccccccc}

				&$(n,m)$ & $(n',m')$&  $n_k$(1QL)&  $n_k$(3QL) \\
				\hline

0.0\degree 	   &( 0 2 )	 &( 0 1 )    & 15 & -       \\
0.0\degree 	 &( 0 5 )	 &( 0 3 )      & 6 & -        \\ 
4.3\degree 	 &( 2 3 )	 &( 1 2 )      & 9 & 3        \\ 
4.7\degree 	 &( 4 3 )	 &( 2 2 )      & 3 & -       \\
8.9 \degree	 &( 1 5 )	 &( 0 3 )      & 3 & -        \\ 10.9\degree 	 &( 2 1 )	 &( 1 1 )      & 15& 9         \\ 
13.9\degree 	 &( 1 3 )	 &( 0 2 )      & 15 & 9       \\ 
16.1\degree 	 &( 3 1 )	 &( 1 1 )      & 15 & -        \\ 
17.5\degree 	 &( 3 2 )	 &( 1 2 )      & 9 & 3       \\
19.1\degree 	 &( 4 0 )	 &( 2 1 )      & 12 & -      \\ 
19.1\degree 	 &( 5 0 )	 &( 2 1 )      & 6 & -        \\ 
19.1\degree     &( 2 4 )       &( 0 3 )      & 3 & -     \\
19.1\degree     &( 2 4 )        &( 0 3 )      & 3 & -     \\
20.8\degree     &( 4 3 )        &( 1 3 )      & 3 & -    \\
21.1\degree     &( 5 1 )        &( 2 2 )      & 3 & -      \\
21.8\degree 	 &( 4 2 )	 &( 1 2 )      & 6 & -        \\ 30.0\degree 	 &( 1 1 )	 &( 0 1 )      & 45 & 45        \\
30.0\degree     &( 7 0 )        &( 2 2 )      & 3 & -    \\

     	\end{tabular}
    \end{ruledtabular}

    \end{table}
%------------------------------------------------------------------------	

%------------------------------------------------------------
%------------------------------------------------------------
\section{Distinguishing the two types of band pairs for $\Theta=30\degree$}
\label{App:finestruc}
\begin{figure*}[htb]
     \includegraphics[width=.99\linewidth]{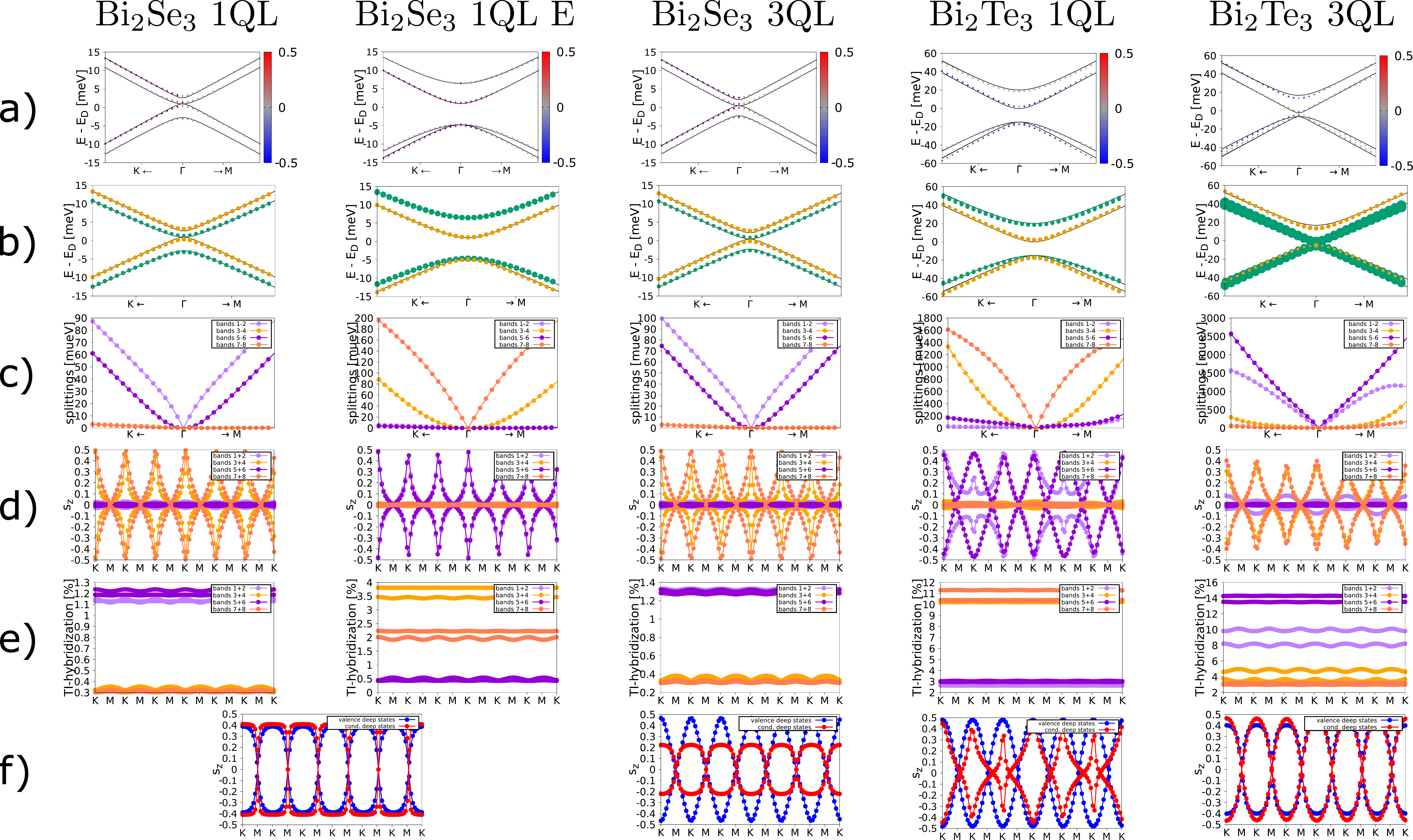} 
     \caption{The two types of band pairs of the 8 band Dirac cone for $\Theta=30\degree$. For selected cases we show the Dirac cone and the distinct properties of the two types of band pairs it consists of. All cases are without applied electric field except 'Bi$_2$Se$_3$ 1Q E', where an electric field of $4.11\frac{V}{nm}$ was applied. a) and b) show the Dirac cone, with color coded spin-z (a) and color coded orbital decomposition (b). In b) green dots show TI s-orbital contribution, while yellow dots shows $m_j=\pm 3 / 2$ orbital contribution as in Fig.~\ref{Fig:ResII}. Line c) shows the ($\mu$eV) splittings of the band pairs, where 'bands 1-2' refers to the energetically lowest pair of bands. All following lines (d)-f)) are concerned with the properties along a circular path around the Dirac cone at $55$~meV.  d) and e) show spin-z expectation values and the contribution of the TI orbitals along this circular path respectively. f) shows (like d)) the spin-z along the circular path, but not for the Dirac cone, but for selected deep lying TI states with major $m_j=\pm  3 / 2$ contribution (see Fig.~\ref{Fig:ResII}). The labels '$K$' and '$M$' in d)-f) indicate that the $\mathbf{k}$-point is lying on the $\Gamma$-$K$ (or the $\Gamma$-$M$) connection line.\label{Fig:finestruc}}
    \end{figure*}

In the main manuscript we touched on the two different kind of band pairs of the Dirac cones in the case of the $\Theta=30\degree$ supercell band structures. We now explore this in detail. In the following we will call band pairs 'type one', if they have similar properties as the energetically lowest pair of bands in the Bi$_2$Se$_3$ 1QL case without electric field (Fig.~\ref{Fig:finestruc},first column). Accordingly, 'type two' band pairs then are similar to the band pair energetically above this 'type one' case.
Fig.~\ref{Fig:finestruc} summarizes the relevant differences of the two types. From Fig.~\ref{Fig:finestruc} a) and d) we see that the spin-z expectation values of 'type one' bands are (almost) zero: both bands are unpolarized in the z-direction. In contrast, 'type two' band pairs show an anisotropic spin structure, where the bands show opposite spin-z values along the $\Gamma$-$K$ line and vanishing values along the $\Gamma$-$M$ line. Additionally, the in-plane spin structures we see in Fig.~\ref{Fig:GlobalBS} are also bound to the type of band pair. Fig.~\ref{Fig:finestruc} b) shows that 'type one' band pairs have relevant contributions from TI $m_j=\pm  3 / 2$ states, while 'type two' band pairs have relevant contributions from TI s-orbital contributions. In Fig.~\ref{Fig:finestruc} c), we show the splittings within the band pairs: Although both 'type one' and 'type two' bands are only split on the $\mu$eV scale, the splittings of 'type one' bands are significantly higher and increase to the meV range, when moving away from the center of the Dirac cone at $\Gamma$. This splitting is enough to overcome the small spin splitting, seen in 'type two' bands. In Fig.~\ref{Fig:finestruc} e), one can see that the TI content of 'type one' band pairs is higher than that of 'type two' band pairs by a factor of about 5 on average. Finally, in Fig.~\ref{Fig:finestruc} f) we see that the same spin structure seen in 'type two' bands can be seen in deep lying (so energetically low valence or energetically high conduction) bands of the TI.
We conclude that on the one hand 'type one' bands acquire their proximity SOC almost exclusively from deep lying TI states imprinting their spin-z structure and their $m_j=\pm 3 / 2$ character. 'type two' band pairs on the hand acquire the majority of their  proximity SOC from the near surface state, which results in a larger splitting destroying the spin-z structure and a significant TI s-orbital content.
Comparing the different cases, we see that different structures can entail a different ordering of 'type one' and 'type two' bands. Shifting the Dirac cone within the TI band structure can result in a switch of this ordering as well. In fact, we observe such a switching of the ordering by electric field for all of the cases in Fig.~\ref{Fig:ResII}, except 3QL of Bi$_2$Se$_3$.
%------------------------------------------------------------------------
%------------------------------------------------------------------------
\section{Alternative fitting Hamiltonian for $\Theta=30\degree$}
\label{App:Aaronfitting}
\begin{figure}[htb]
     \includegraphics[width=.99\linewidth]{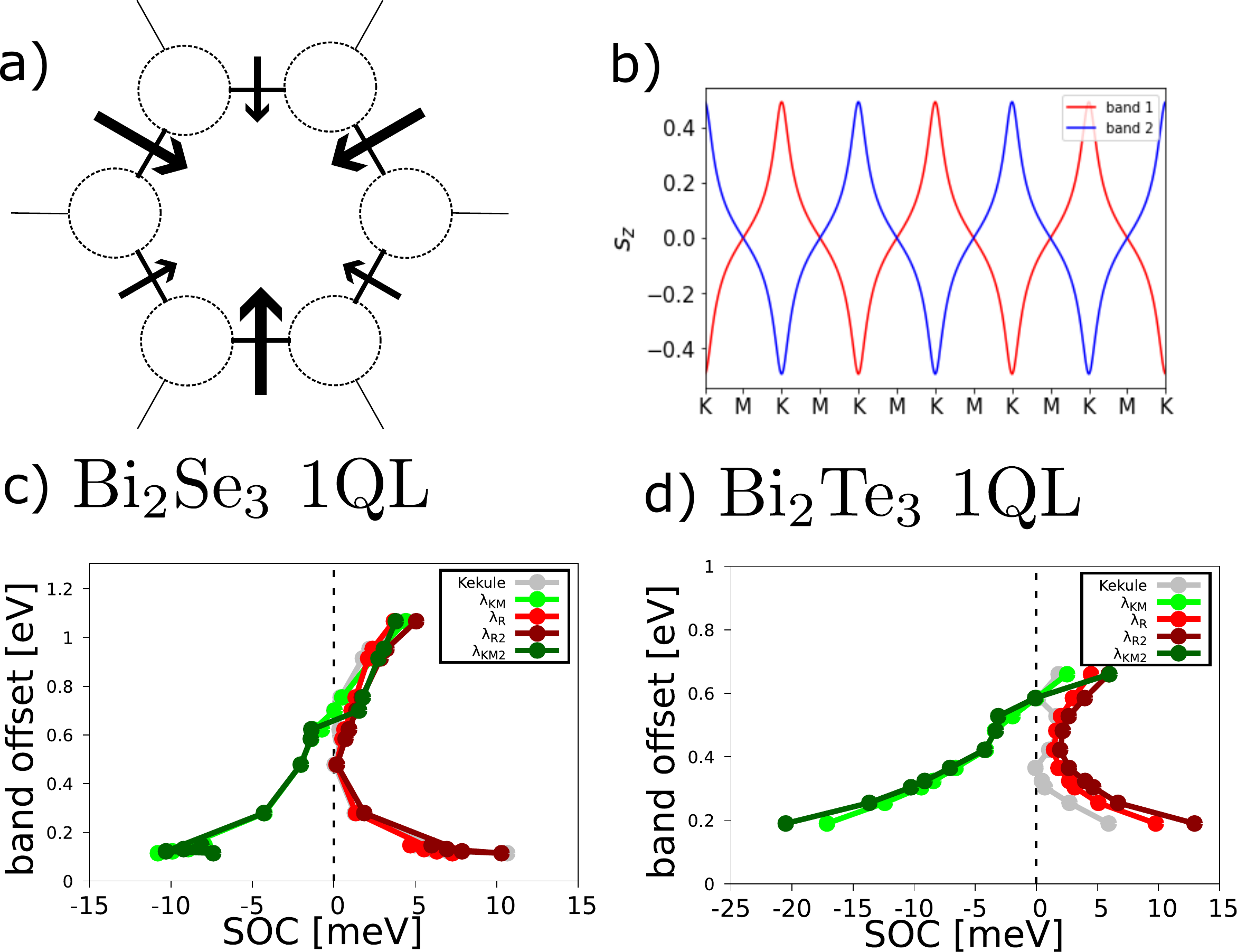} 
     \caption{
    Alternative fitting results: a) shows a schematic of the additional fitting parameters $\lambda_{R,2}$ and $t_0-t_p$ introducing an in-plane Rashba effect and a Kekule distortion  effect. The carbon ring beneath the Se (or Te) atom closest to the graphene plane is shown. The electric field producing the in-plane Rashba effect is depicted by arrows and the electric field along the smaller arrows is set to be only 20\% compared to the one along the larger arrows. The thicker connection lines between the carbon atoms indicate increased hoppings by the Kekule distortion. b) shows an exemplary (Bi$_2$Se$_3$ 1Q, without electric field correction) spin-z structure along a circle at $\approx 60~meV$ above the Dirac cone. The labels '$K$' and '$M$' indicate that the $\mathbf{k}$-point is lying on the $\Gamma$-$K$ (or the $\Gamma$-$M$) connection line. c) and d) show the SOC parameters $|\lambda_R|$ and $\lambda_{KM}$ (extracted with Eq.~\eqref{Eq:Ham}, see Fig.~\ref{Fig:ResII}) in comparison with the parameters $\lambda_{R,2}$, $\lambda_{KM,2}$ and $t_0-t_p$ (extracted with Eq.~\eqref{Eq:Ham2}) for different electric fields. The electric field ranges used to shift the graphene Dirac cone within the TI bands is the same as in Fig.~\ref{Fig:ResII}.
     }\label{Fig:Aaronfitting}
    \end{figure}

The Dirac cone of the 'Hollow' $\Theta=30\degree$ supercell has an intricate spin structure (see Fig.~\ref{Fig:finestruc}). In the main manuscript we neglect this fine structure, since it is based on $\mu$eV splittings within the band pairs. Now, we try to describe this fine structure using a full tight binding Hamiltonian akin to the one used in Ref.~\citep{Song2018:TIGRheteroDFT}:
\begin{equation}
\begin{split}
    H=&\sum_{\alpha=0,p} t_{\alpha}\sum_{\langle ij\rangle,s} c_{is}^\dagger c_{js} + \sum_{i,s}\Delta_i c_{is}^\dagger c_{is}\\
    &+\frac{i}{3\sqrt{3}}\sum_{\langle\langle ij\rangle\rangle,ss'}c_{is}^\dagger c_{js'}(\lambda_{KM,2}+\xi \lambda_{VZ})[\nu_{ij}s_z]_{ss'}\\
    &+\frac{2i}{3}\sum_{\langle ij\rangle,ss'}c_{is}^\dagger c_{js'} [(\lambda_R\mathbf{\hat{z}}+\lambda_{R,2}\bm{\rho})\cdot(\mathbf{s}\times\mathbf{d}_{ij})]_{ss'} 
\end{split}
\label{Eq:Ham2}
\end{equation}
The single brackets constitute sums over nearest neighbours and the double brackets constitute sums over next nearest neighbours. $c^\dagger_{is}$ and $c_{is}$ are the creation and annihilation operators of an electron at site $i$ with spin $s$, $\mathbf{d}_{i,j}$ is a unit vector pointing from site j to nearest neighbour site i, $\mathbf{s}$ is a vector containing the Pauli matrices, $\nu_{ij}$ is equal to +1 for clockwise and equal to -1 for counterclockwise hoppings from site j to i, $\xi$ is +1 for sublattice A and -1 for sublattice B, $\mathbf{\hat{z}}$ is the unit vector in $z$-direction and $\bm{\rho}$ is an in-plane vector representing the electric fields in Fig.~\ref{Fig:Aaronfitting}~a). The first term of $H$ describes the orbital part with $t_0$ as (stronger) hopping within the carbon ring depicted in Fig.~\ref{Fig:Aaronfitting}~a) and $t_p$ as the (weaker) hopping parameter between such carbon rings. This makes $t_0-t_p$ the Kekule distortion parameter. The second term describes a series of onsite potentials $\Delta_i$, which are only used in the fitting of the 'Top' case. The third term describes Kane-Mele and valley-Zeeman SOC. Note that this term is an exact translation of the Kane-Mele and valley-Zeeman terms in Eq.~\eqref{Eq:HamKMVZ} in contrast to Ref.~\citep{Song2018:TIGRheteroDFT}, where the Kane-Mele hoppings only exist on the central carbon ring. The fourth term describes a Rashba SOC related to an electric field in $z$-direction (again an exact translation of the term in Eq.~\eqref{Eq:HamR}) and an additional Rashba term $\lambda_{R,2}$ related to electric fields $\bm{\rho}$ depicted as arrows in Fig.~\ref{Fig:Aaronfitting}~a). The electric fields are all pointing inward to the carbon ring located under the  Se (or Te) atom closest to graphene. There are two distinct types of arrows: to one  we assign  $|\bm{\rho}|=1$ (large arrows in Fig.~\ref{Fig:Aaronfitting}~a)), to the other $|\bm{\rho}|=0.2$ (small arrows in Fig.~\ref{Fig:Aaronfitting}~a)). Note that Ref.~\citep{Song2018:TIGRheteroDFT} do not allow for the in-plane Rashba connected to the smaller arrows.

For the 'Hollow' case as in the manuscript we have $\lambda_{VZ}=\Delta_i=0$. We obtain good fits only for very small, but non-zero values for $|\lambda_R|$ and therefore simply fix it to $|\lambda_R|=0.01$~meV for all fits. The remaining parameters therefore are: $\lambda_{KM,2}$, $t_0-t_p$ and $\lambda_{R,2}$. The general spin-z structure of the 'type two' $\Theta=30\degree$ band pairs can be modeled with the fittings (see Fig.~\ref{Fig:Aaronfitting}~b)). To model the spin-z structure of a 'type one' band pair, one can simply set the magnitude of all in-plane electric fields to equal values $|\bm{\rho}|=0.6$ for the directions indicated by both the large and the small arrows. This will not significantly change the values of the other fitting parameters. Note that the in-plane spin structure also cannot be sufficiently reproduced using this Hamiltonian.

We show the fittings results in Fig.~\ref{Fig:Aaronfitting}~c) and d) for Bi$_2$Se$_3$ and Bi$_2$Te$_3$ respectively for the same electric fields as in Fig~\ref{Fig:ResII}. We see  the same results for Kane-Mele SOC (compare light green to dark green curve). Also the new in-plane Rashba SOC replaces the  out-of-plane Rashba SOC (compare red to dark-red curve). The Kekule distortion parameter will follow a similar behaviour as the Rashba SOC increasing drastically in the vicinity of a TI band.

%------------------------------------------------------------------------
%------------------------------------------------------------------------
%------------------------------------------------------------
\section{Effect of lateral shifting}
\label{App:shift}
For incommensurate heterostructures the lateral shifting degree of freedom does not play a role. Assuming a sample infinite in $x$- and $y$-direction, every shifting configuration exists somewhere on the sample. The physical properties of the configurations will then average out, when considering the properties of the whole material. However, the structures used in our DFT calculations are commensurate to be
computationally viable and the lattice constants are forced by strain to be compatible. Therefore, different lateral relative shifts might ensue different physical properties including different proximity induced SOC.

Naturally, this effect is less relevant for large supercells, for which an averaging over the different shifts will occur within the supercell. We show this by investigating the shifting-dependence for two supercells: the 13.9\degree~and the 30\degree~supercell. While the former consists of a total of 46 atoms and gives enough area for the different configurations to average out, the latter only consists of 13 atoms. Therefore, different lateral shifting positions will lead to different effects on the Dirac cone for the 30\degree~supercell.

\begin{figure}[htb]
     \includegraphics[width=.99\linewidth]{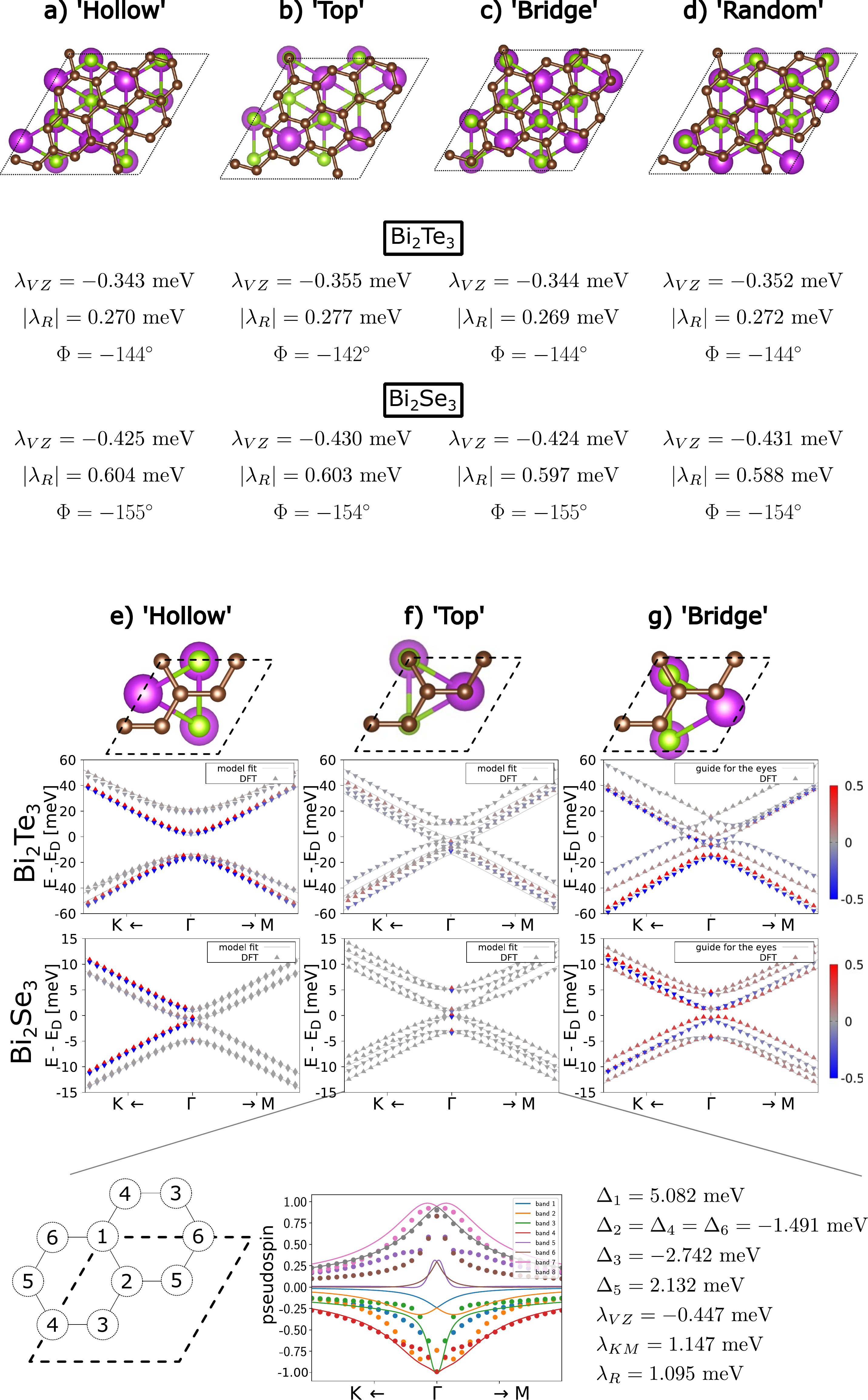} 
     \caption{
    We show supercells with different lateral shifting positions for the $\Theta=13.9\degree$ case (a)-d)) and the $\Theta=30\degree$ case (e)-g)). For the former, we list the relevant extracted SOC parameters $|\lambda_R|$, $\lambda_{VZ}$ and $\Phi$ for both Gr/Bi$_2$Te$_3$ and Gr/Bi$_2$Se$_3$. For the latter we show the proximity SOC modified Dirac cone band structure with color coded spin-z. The triangles represent DFT data. For the cases where satisfactory fittings were possible (e) and f)) the grey lines represent the energies of the fit, while for g) the grey lines represent only a guide to the eyes connecting the data points of the same band. For the 'Top' case of Gr/Bi$_2$Se$_3$ we additionally present the fitting results of the pseudospin and the extracted fitting parameters. The sketch on the left shows the labeling of the atoms in the 'Top' case.
     }\label{Fig:shift}
    \end{figure}

We defined 4 different shifting options, 'Hollow', 'Top', 'Bridge' and 'Random', which we show in Fig.~\ref{Fig:shift}. The keywords indicate how the Te atom closest to the graphene layer is positioned with respect to the graphene structure. For the 30\degree~case there is only one such atom per unit cell, while for the 13.9\degree~supercell there are multiple ones, with different positions (we describe the one closest to the corner of the supercell). Therefore, in the 13.9\degree~case, the Dirac cones for the different shifting positions look almost the same, as the extracted fitting parameters show hardly any difference. In contrast, the Dirac cone of the 30\degree~case varies widely depending on the lateral shift.

We now focus on the 30\degree~supercell. Since the 'Hollow' configuration is the energetically most favorable and has the simplest band structure, we focus on it in the main manuscript. Additionally, we fitted the Dirac cone of the 'Top' case to a full tight binding Hamiltonian, since the simple model Hamiltonian (Eq.~\eqref{Eq:Ham}) in the main manuscript is not able to capture the features of the band structure, which come from the specific shifting position with respect to the $\sqrt{3}\times\sqrt{3}$ graphene supercell. By using the same SOC parameters $\lambda_{VZ}$, $\lambda_{KM}$ and $|\lambda_R|$ and additionally extending the sublattice imbalance $\Delta$ to a general onsite potential $\Vec{\Delta}$ (see Eq.~\eqref{Eq:Ham2}) for the six carbon atoms, we were able to describe the Dirac cone of the 'Top' configuration reasonably well (see Fig.~\ref{Fig:shift}). For all eight bands, the spin-z expectation values are zero, while the pseudospin shows an intricate structure. For the 'Bridge' case, we were not able to find satisfying fitting parameters.
%------------------------------------------------------------------------
%------------------------------------------------------------------------
%------------------------------------------------------------

%------------------------------------------------------------
%------------------------------------------------------------
\section{Effect of increased thickness of the TI slab}
\label{App:QL}
Varying the thickness of the TI layer changes its band structure. When increasing the number of Quintuple layers (QLs), one can see the TI surface state at $\Gamma$ fully forming. While for 1 QL the overlap between the upper and lower surface states opens a large gap, the surface states of the 3QL case are spatially separated enough for the characteristic linear sloped surface state to arise. Since the graphene
 Dirac cone of the $\Theta=30\degree$ supercell interacts directly with the TI surface state, it will be affected significantly by the exact form of the surface state and therefore by the thickness of the TI. 
 
 But also, for the other twist angles, a dependency on the TI thickness is not excluded, since also the TI bands away from $\Gamma$ are influenced by this change which can in turn then change the proximity SOC. To investigate this, we calculated the band structures for the 3QL cases of some of our supercells and gather the extracted parameters in Tab.~\ref{Tab:QL}. Note, that we compare the results without electric field corrected band offsets, because the 3QL TI band structure is significantly more susceptible to unwanted distortion by the electric field. However, since we are only comparing the 1QL case with the 3QL case, which both have the same strain, using the parameters from the uncorrected calculations is a valid approach. The table shows, that while in most of the scenarios the SOC is not strongly affected by the TI thickness, there are two cases highlighting the importance of the TI thickness:
 \begin{enumerate}
     \item For the Gr/Bi$_2$Te$_3$ $\Theta=4.3\degree$ supercell, the change from 1QL to 3QL changes the Rashba phase angle by a significant amount from $152\degree$ to only $79\degree$.
     \item  For the Gr/Bi$_2$Se$_3$ $\Theta=17.5\degree$ supercell, the 1QL case is dominated by valley-Zeeman type SOC $\lambda_{VZ}$, while in the 3QL case the Rashba type SOC $|\lambda_R|$ is dominating. Here, the vicinity to the TI surface state in $k$ space might already be a reason for this massive change in proximity SOC.
 \end{enumerate}
In conclusion, we see that for $0\degree <\Theta \lessapprox 20\degree$ the proximity SOC dependency on the TI thickness is overall not to large, but should not fully be neglected.

\begin{table}[]
\caption{Comparison between extracted SOC parameters for selected cases between structures with 1QL thick TI slabs and 3QL thick TI slab. The parameters are extracted from calculations without electric field corrections.
\label{Tab:QL}}
\begin{tabular}{c|c|cccc}

thickness&$\Theta$[\degree]&$\lambda_{KM}$[meV]&$\lambda_{VZ}$[meV]&$|\lambda_{\text{R}}|$[meV]&$\Phi$ [\degree]\\
\hline
\hline
\multicolumn{4}{l}{Bi$_2$Te$_3$}\\
\hline
1QL&    4.3& -0.01 & 2.04  &0.52        &152\\
3QL&    4.3 & 0.01 & 1.65 &0.50       &79\\

\hline
1QL&    13.9& 0.01 & -0.46 &0.18 & -130 \\
3QL&    13.9& 0.01 & -0.36 &0.28  & -142\\

\hline
1QL&    17.5& 0.16 & 3.51&2.90  & -10 \\
3QL&    17.5& 0.08 & 3.35 &2.06 &-25 \\

\hline
\multicolumn{4}{l}{Bi$_2$Se$_3$}\\
\hline
1QL&    4.3& -0.01 & 1.19 &0.62  & -178  \\
3QL&    4.3& 0.00 & 1.10  &0.38 &171 \\

\hline
1QL&    10.9& 0.01 & 3.18 &1.02  & 47\\
3QL&    10.9& 0.01 & 3.24 &1.00 & 46 \\

\hline
1QL&    13.9& 0.01 & -0.50 &0.62 &-155 \\
3QL&    13.9& 0.00 & -0.43 &0.60 &-154\\

\hline
1QL&    17.5& 0.03 & -0.55 &0.08  & 17 \\
3QL&    17.5& -0.00 & 0.02 &0.32 &23\\
\hline
\end{tabular}

\end{table}

%------------------------------------------------------------

%\bibliographystyle{naturemag}
\bibliography{references}

\end{document}